\documentclass[aps,prb,showpacs,letterpaper,preprint]{revtex4}
\usepackage{graphicx}
\usepackage{verbatim}
\usepackage{amsfonts,amsmath,amssymb,bm}

\begin{document}

\title{Sum-over-states vs quasiparticle pictures of
coherent correlation spectroscopy of excitons in 
semiconductors; femtosecond analogues of multidimensional NMR
}
\date{\today{}}
\author{Shaul Mukamel, Rafal Oszwaldowski, Darius Abramavicius}
\affiliation{Chemistry Department, University of California, Irvine, CA 92697-2025,
United States}

\begin{abstract}
Two-dimensional correlation spectroscopy (2DCS) based on the nonlinear
optical response of excitons to sequences of ultrafast pulses, has the
potential to provide some unique insights into carrier dynamics in
semiconductors. The most prominent feature of 2DCS, cross peaks, can best be
understood using a sum-over-states picture involving the many-body
eigenstates. However, the optical response of semiconductors is usually
calculated by solving truncated equations of motion for dynamical variables,
which result in a quasiparticle picture. In this work we derive Green's
function expressions for the four wave mixing signals generated in various
phase-matching directions and use them to establish the connection between
the two pictures. The formal connection
with Frenkel excitons (hard-core bosons) and vibrational excitons (soft-core
bosons) is pointed out.
\end{abstract}

\maketitle

\section{Introduction}

Exciton models are widely used to describe the linear and nonlinear optical
properties of many types of systems, including semiconductor nanostructures
(quantum wells, dots and wires), molecular aggregates and crystals,\cite%
{Davydov1962,Mukamel_inzyss1993,ChernyakZhangMukamel98} as well as
vibrations in proteins.\cite{ZhuangMukamel2005,MukamelAbramavicius2004} In
semiconductors, nonlinear optical experiments reveal a wealth of interesting
phenomena.\cite{ChemlaShah2001,ElsaesserPRL2001,Shih2005,Shahbook,Meierbook} 
For
instance, such experiments provide information about many-exciton states
such as biexcitons, their interactions, relaxation and dissociation.\cite%
{Chemla2001_semi_rev,AdachiLitton2004,RiceYasin2005,DanckwertsKnorr2006}

The introduction of multidimensional techniques had revolutionized NMR in the
seventies\cite{AueW1976} and established it as a powerful tool for studying
complex
systems and identifying specific structural and dynamical correlations.\cite%
{Ernstbook} In such experiments the system is subjected to a sequence of
well separated pulses.
Correlation plots of the signals vs. two (or more) time delay periods then
provide multidimensional spectroscopic windows into the system. The correlated
dynamics of
spins carefully prepared by the pulse sequence is very sensitive to their
interactions. Analysis of these correlation plots then provides a powerful
probe for molecular geometries and dynamical correlations.
 These techniques were recently extended to the infrared and
the visible regime and were shown to be very useful for Frenkel excitons in
molecular systems.\cite%
{mukamel_annrev2000,TaniMuk1993,MukamelAbramavicius2004,
ZhangMukamel1999} There are some differences between the optical and the NMR
techniques. NMR uses strong saturating fields whereas optical pulses are most
effective in the weak field regime. NMR signals are essentially isotropic in
space
whereas coherent optical signals are generated in well defined (phase-matching)
directions. These differences were explored in detail in 
Refs.~\onlinecite{Scheurer2002,Sche2001,Sche2002}. Nevertheless
the NMR and optical techniques are conceptually similar and many ideas of pulse
sequences  developed in NMR
may be adopted in the optical regime, where the millisecond NMR time-scale is
pushed to the femtosecond regime.
The same ideas may be extended 
to study interband and intersuband excitations in semiconductors.\cite%
{Lijun2006,Kuzn2007,Erem2006,CundiffPRL2006,Borc2005}
Multidimensional analysis of the nonlinear optical response of
semiconductors to sequences of femtosecond pulses could provide a novel
probe for many-body interactions. In a recent work\cite{Lijun2006} 
on semiconductor Quantum Wells,
2D correlation spectra from three 3rd order optical techniques have
 been calculated. The unique character of 2D spectroscopy allowed to easily 
recognize and classify features due to different types of biexcitons. Such
features are sometimes difficult to separate in the 
usual one-dimensional mode of displaying non-linear spectra,  due 
to the strong line broadening and the highly congested exciton spectra.

Two types of approaches have been traditionally used
towards modeling the nonlinear optical response of excitonic systems. The
first is based on the many-body eigenstates obtained by exact
diagonalization of the Hamiltonian.\cite{mukbook} Sum-over-states (SOS)
expressions can then be derived for the nonlinear response functions and
optical signals. This method is practical in many applications to electronic
and vibrational Frenkel excitons in molecules\cite%
{Brixner-fleming-nature2005,FangHochstrasser2004,AbramaviciusMukamel2004}
and allows clear identification and classification of possible 
single- and multi-photon
resonances. Calculating the eigenstates is a serious computational
bottleneck in extended structures. For an $N$ site tight-binding
Frenkel-exciton model the number of single and two-exciton states scales as
$\sim N$ and $\sim N^{2}$ respectively. For Wannier excitons in semiconductors
these scalings are $\sim N^{2}$ and $\sim N^{4}$, making the simulations
prohibitively expensive. This is why the approach is not widely used for
electron-hole excitations in semiconductors. Instead, one adopts a second
strategy, which describes the response in terms of quasiparticles (QP), and
the many-particle eigenstates are never calculated. \cite%
{spanomuk89,leegwaterMukamel92,mukbook,ChernyakMukamel1996,spanomuk91a,Mukamel_inzyss1993}
Calculations are performed by solving equations of motion for microscopic
coherences, which are coupled to other dynamical variables. Even for a
simple system such as a single semiconductor quantum well, solving the
equations numerically to create a 2D map of a nonlinear response function is
computationally expensive,\cite{Lijun2006} since these equations must be
solved repeatedly for different pulse delays. Only after obtaining the
optical signal on a 2D time grid, a Fourier transform can be performed to
get the 2DCS.
Apart from direct, numerical solutions of equations of motion
\cite{Koch1999,Weis2000} there exist other theoretical approaches to exciton
correlation effects, such as memory kernel representation
\cite{Ostr1995,Axt2001} or Coupled Cluster Expansion for doped semiconductors.%
\cite{Prim2000,Shah2000}

In this paper we derive closed expressions for 2DCS of semiconductors by
solving the Nonlinear Exciton Equations (NEE)\cite%
{AxtMukamel1998rev,ChernyakZhangMukamel98} for the third order response. 
Both time-ordered and non-ordered forms of the response function which
represent time and frequency domain techniques, respectively, are derived.
Our QP expressions for the response are given in terms of the single exciton
Green's function and the exciton scattering matrix. The SOS response
functions, in contrast, are expressed in terms of many-exciton eigenstates. 
Even
though the response functions calculated using both techniques must be
identical, the relation between the two pictures is not obvious.
The expressions look very
different and it is not possible to see their equivalence by a simple
inspection. The SOS expressions contain large terms, which grow with system
size and have opposite signs, thus they
almost cancel. This complicates their numerical implementation. In contrast
these cancellations are built-in from the outset in the QP approach, which uses a
harmonic reference system. The nonlinearities are then attributed to
exciton-exciton scattering which is absent in the harmonic reference system.
The second goal of this
paper is to show precisely how the two pictures of many-body correlations are
connected. 
We write down the SOS expressions using the Keldysh loop and then derive the QP
expressions directly from the SOS ones. This provides a time-domain
interpretation for the interference effects. 
The SOS and the QP expressions provide complementary views into the origin
of features seen in 2D spectrograms.

In Sec.~\ref{Sec:SOS} we present the SOS expressions for the third order
response obtained from time-dependent perturbation theory. Their QP
counterparts are derived in Sec.~\ref{Sec:QP}. We use the method developed
in Refs.~\onlinecite{ChernyakMukamel1996,ChernyakZhangMukamel98} to
transform the Hamiltonian to a form typical for interacting oscillators. The
starting many-electron Hamiltonian can be written in an \textit{ab-initio},%
\cite{Oszw2005} tight-binding \cite{SiehPRL1999} or a $\mathbf{k}\cdot%
\mathbf{p}$ basis. One of the key results of this paper, i.e., the
equivalence of the SOS and QP pictures is proven in Sec.~\ref{Sec:Connection}%
. In Sec.~\ref{Sec:2D} we derive closed expressions for 2D correlation
signals. 
The QP approach provides a unified description for electron-hole
excitations in semiconductors as well as to Frenkel excitons in molecular
aggregates (Paulions) and anharmonic vibrations (bosons), which are
described by the same general Hamiltonian. QP formulae for nonlinear
response have been derived previously along similar lines for Frenkel
excitons. This connection is shown in Appendix \ref{App_coup_osc}. 
In
the last Section (\ref{Sec:Discuss}) we discuss the results.

\section{Sum-over-states expressions for the time-ordered nonlinear response
\label{Sec:SOS}}

\begin{figure}[p]
\includegraphics[clip=true, viewport=5.0in 2.0in 6.5in 8.0in,scale=1.0,
angle=-90]{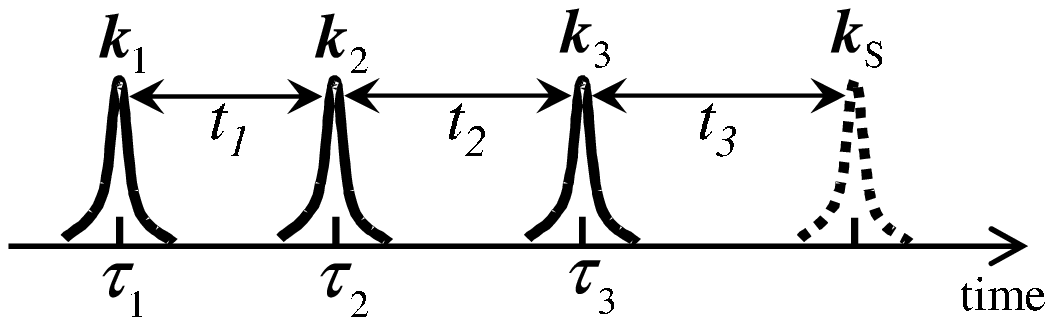} 
\caption{The sequence of light pulses in a time-domain Four Wave Mixing
Experiment: the pulses are centered at times
$\tau_1$, $\tau_2$,
$\tau_3$, while the delays are $t_1$, $t_2$ and $t_3$. (The latter are 
sometimes denoted as $\tau$, $T$ and $t$.) The signal is generated in
the ${\bm k}_S$ direction.}
\label{fig:pulses}
\end{figure}

We consider a 4 wave-mixing experiment performed with three femtosecond laser
pulses (Fig.~\ref{fig:pulses}). The optical electric field is:
\begin{equation}
E\left( \bm{r},t\right) =\sum_{j=1}^{3}E_{j}\left( \bm{r},t\right) =E^{+}(%
\bm{r},t)+E^{-}(\bm{r},t),   \label{eqn:Field}
\end{equation}%
\begin{equation}
E^{+}(\bm{r},t)=\sum_{j=1}^{3}\mathcal{E}_{j}^{\mathcal{+}}(t-\tau
_{j})e^{-i\omega_{j}t}e^{i\bm{k}_{j}\bm{r}},   \label{eqn:FieldA}
\end{equation}%
\begin{equation}
E^{-}(\bm{r},t)=\sum_{j=1}^{3}\mathcal{E}_{j}^{-}(t-\tau
_{j})e^{i\omega_{j}t}e^{-i\bm{k}_{j}\bm{r}}.   \label{eqn:FieldB}
\end{equation}
The $j$-th pulse is centered at $\tau_{j}$, has an envelope $\mathcal{E}%
_{j}(t-\tau_{j})$, carrier frequency $\omega_{j}$, and wavevector $\mathbf{k}%
_{j}$. $E^{+}$ ($E^{-}$) denotes the positive (negative) frequency part of
the field, and $\mathcal{E}_{j}^{-}=\left( \mathcal{E}_{j}^{+}\right) ^{\ast}
$. The induced polarization in the system is recorded as a function of
time-delays between pulses.%

Assuming the dipole interaction with the optical field $\hat{H}_{I}={\hat{\bm%
\mu}}\cdot\bm{E}(\bm{r},\tau)$, where ${\hat{\bm\mu}}$ is the dipole
operator, the third-order contribution to the system's polarization can be
written as 
\begin{equation}
\bm{P}(\bm{r},\tau_{4})=\iiint_{-\infty}^{\infty}\mathrm{d}\tau_{3}\mathrm{d}%
\tau_{2}\mathrm{d}\tau_{1}\bm{S}^{(SOS)}(\tau_{4},\tau_{3},\tau_{2},\tau_{1})%
\bm{E}(\bm{r},\tau_{3})\bm{E}(\bm{r},\tau_{2})\bm{E}(\bm{r},\tau_{1}), 
\label{eqn:Polarization}
\end{equation}
where the response function $\bm{S}^{(SOS)}$, which connects the induced
polarization with the laser field envelopes, is given by (throughout this
paper we set $\hbar=1$): 
\begin{align}
\bm{S}^{(SOS)}(\tau_{4},\tau_{3},\tau_{2},\tau_{1}) & =i^{3}\left[
\theta(\tau_{43})\theta(\tau_{32})\theta(\tau_{21})\left\langle \hat{\bm{\mu}%
}(\tau_{4})\hat{\bm{\mu}}(\tau_{3})\hat{\bm{\mu}}(\tau_{2})\hat{\bm{\mu}}%
(\tau_{1})\right\rangle \right.  \label{eqn:Response} \\
& -\theta(\tau_{43})\theta(\tau_{42})\theta(\tau_{21})\left\langle \hat{%
\bm{\mu}}(\tau_{3})\hat{\bm{\mu}}(\tau_{4})\hat{\bm{\mu}}(\tau_{2})\hat{%
\bm{\mu}}(\tau_{1})\right\rangle  \notag \\
& +\theta(\tau_{42})\theta(\tau_{23})\theta(\tau_{41})\left\langle \hat{%
\bm{\mu}}(\tau_{3})\hat{\bm{\mu}}(\tau_{2})\hat{\bm{\mu}}(\tau_{4})\hat{%
\bm{\mu}}(\tau_{1})\right\rangle  \notag \\
& \left. -\theta(\tau_{41})\theta(\tau_{12})\theta(\tau_{23})\left\langle 
\hat{\bm{\mu}}(\tau_{3})\hat{\bm{\mu}}(\tau_{2})\hat{\bm{\mu}}(\tau_{1})\hat{%
\bm{\mu}}(\tau_{4})\right\rangle \right] .  \notag
\end{align}
We shall use double-sided Feynman diagrams to represent the
time ordering of various interactions.\cite{MukaX2005}
The four terms in Eq.~(\ref{eqn:Response}) are represented by diagrams
a, b, c, d shown on Figure \ref%
{fig:nrf3-partially-ordered}. These diagrams should be read starting at the
bottom left and proceeding along the loop, clockwise, as indicated by the
arrows. The $\tau_{i}$ variables are ordered on the Keldysh-Schwinger loop,
but not necessarily in real (physical) time. $\tau_{i}$ in diagrams (a) and
(d) are also ordered in real time. This is not the case for diagrams~(b) and
(c): in (b) $\tau_{3}$ can come either before or after $\tau_{1}$ and $%
\tau_{2}$, whereas in (c) $\tau_{1}$ can come either before or after $\tau_{3}$ and $%
\tau_{2}$. 
\begin{figure}[p]
\includegraphics[clip=true, viewport=1.0in 1.0in 5.0in 9.0in,scale=0.5,
angle=-90]{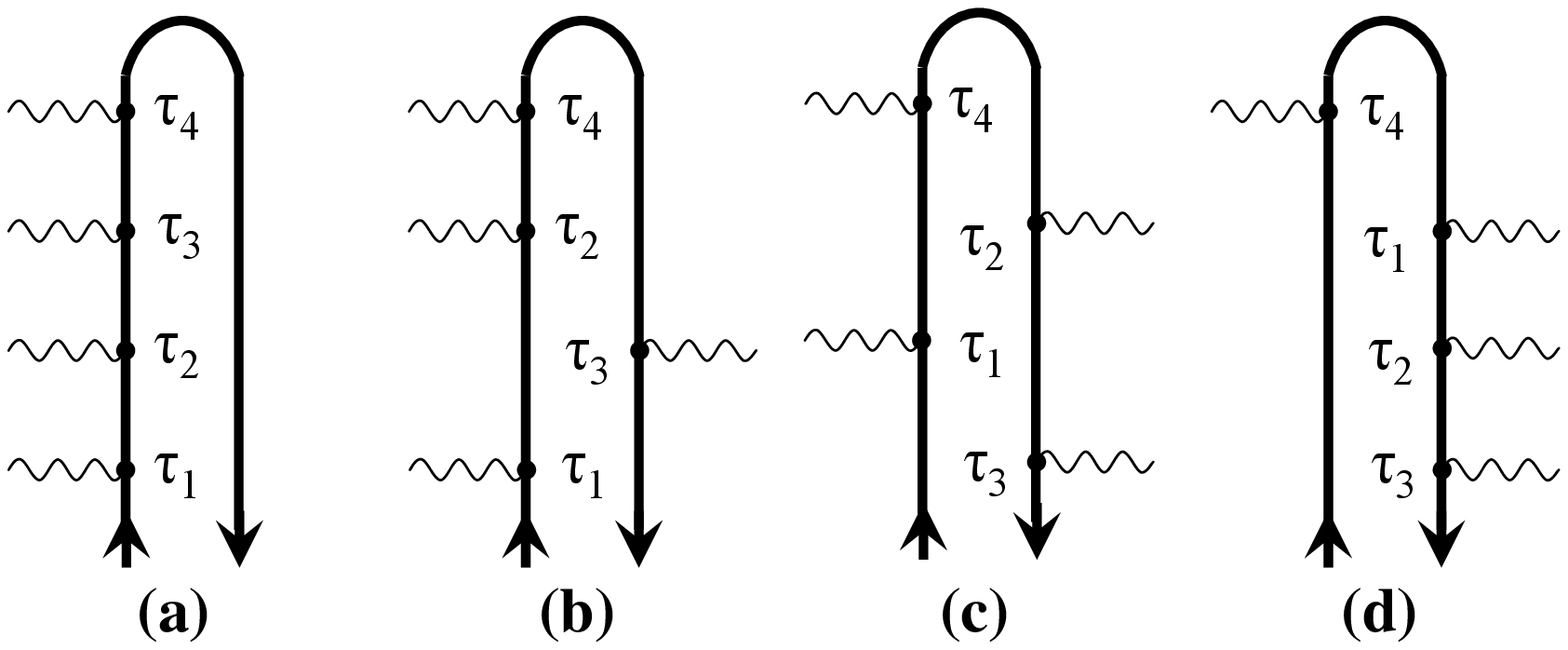}
\caption{Diagrams representing the four partially time-ordered terms
contributing to the third-order polarization (Eq.~\protect\ref{eqn:Response}%
).}
\label{fig:nrf3-partially-ordered}
\end{figure}

If the eigenstates $|a\rangle$ and eigenvalues $\varepsilon_{a}$ of the
system are known, Eq.~(\ref{eqn:Response}) may be expanded in terms of the
corresponding matrix elements:%
\begin{align}
& \left\langle \bm{\mu}(\tau_{4})\bm{\mu}(\tau_{3})\bm{\mu}(\tau_{2})\bm{\mu}%
(\tau_{1})\right\rangle  \label{eqn:cf4sos} \\
& =\sum_{a_{1},a_{2},a_{3}}\bm{\mu}_{ga_{3}}\bm{\mu}_{a_{3}a_{2}}\bm{\mu}%
_{a_{2}a_{1}}\bm{\mu}_{a_{1}g}e^{-i\big[(\varepsilon_{a_{3}}-%
\varepsilon_{g})\tau_{4}+(\varepsilon_{a_{2}}-\varepsilon_{a_{3}})\tau_{3}+(%
\varepsilon_{a_{1}}-\varepsilon_{a_{2}})\tau_{2}+(\varepsilon_{g}-%
\varepsilon_{a_{1}})\tau_{1}\big]}~.  \notag
\end{align}

\begin{figure}[p]
\begin{center}
\includegraphics[width=0.8in]{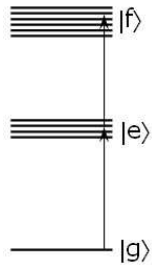}
\end{center}
\caption{Energy levels of the exciton model.}
\label{fig:exciton-model}
\end{figure}

So far we considered a general multilevel system. 
We next turn to the response of excitons, where the energy levels form manifolds,
classified by the number of excitons: the ground state $\left( g\right) $,
single exciton $\left( e\right) $, two-exciton $\left( f\right) $ (or
biexciton), etc. (Fig. \ref{fig:exciton-model}). We shall assume that the
dipole operator can only create and annihilate a single exciton at a time.
Only the single and the two-exciton states then contribute to the third
order signals. We further partition the dipole operator as $\hat{\bm{\mu}}=%
\hat{\bm{\mu}}^{+}+\hat{\bm{\mu}}^{-}$, where $\hat{\bm{\mu}}^{+}$ is the
positive frequency part which induces upward $g$ to $e$ and $e$ to $f$
transitions, while its Hermitian conjugate $\hat{\bm{\mu}}^{-}$ (the
negative frequency part) induces the opposite transitions. We thus write%
\begin{align*}
\hat{\bm{\mu}}^{+} & =\sum_{\varepsilon_{\nu}>\varepsilon _{\nu^{\prime}}}%
\bm{\mu}_{\nu\nu^{\prime}}\left\vert \nu\right\rangle \left\langle
\nu^{\prime}\right\vert , \\
\hat{\bm{\mu}}^{-} & =\sum_{\varepsilon_{\nu}<\varepsilon _{\nu^{\prime}}}%
\bm{\mu}_{\nu\nu^{\prime}}\left\vert \nu\right\rangle \left\langle
\nu^{\prime}\right\vert .
\end{align*}
Invoking the rotating-wave approximation (RWA), we neglect all terms where
at least one of the transitions is not in resonance with one of the incident
carrier frequencies. The system-field interaction term then becomes%
\begin{equation*}
H_{I}\left( t\right) =-\hat{\bm{\mu}}^{+}\bm{E}^{+}(\bm{r},\tau)-\hat{%
\bm{\mu}}^{-}\bm{E}^{-}(\bm{r},\tau) 
\end{equation*}

\begin{figure}[ptb]
\begin{center}
\includegraphics[clip=true, viewport=1.0in 4in 8.0in 7in,scale=0.5]{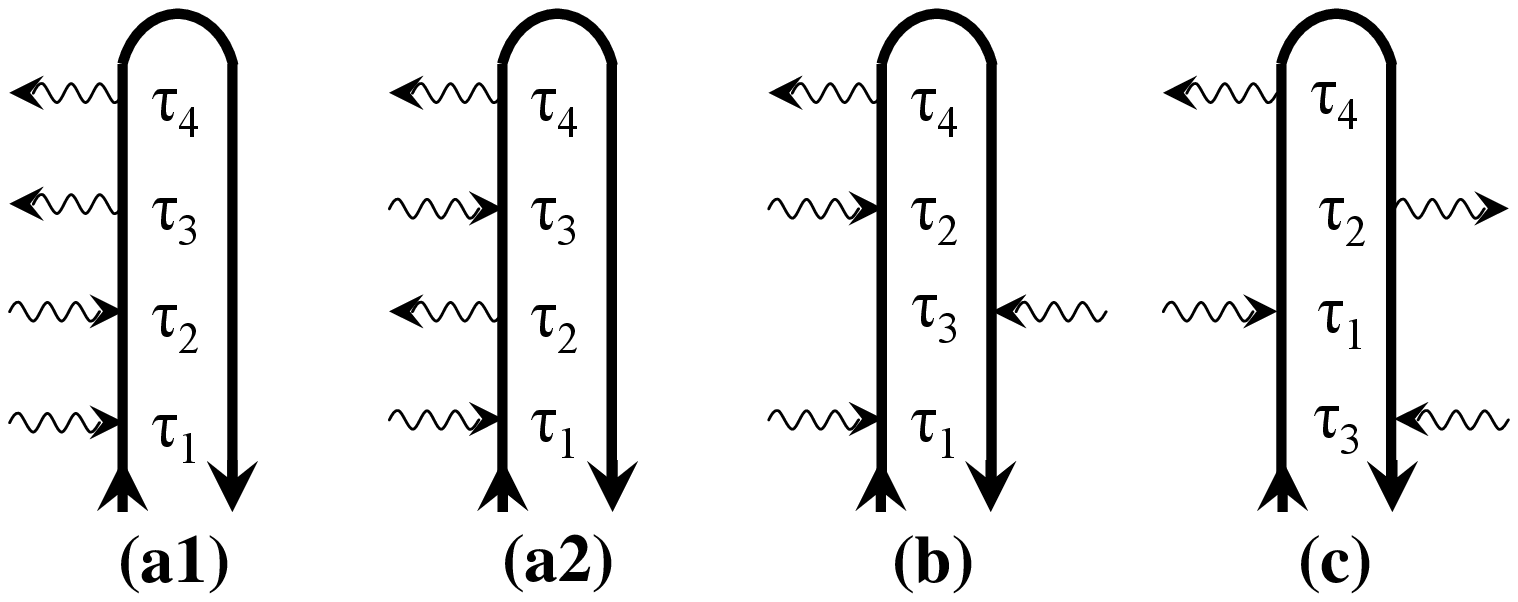}
\end{center}
\caption{Loop diagrams representing the partially time-ordered terms
contributing
to the third-order polarization [Eq.~(\protect\ref{eqn:ResponseA})] within
the rotating wave approximation. Arrows pointing to the right represent $
\bm{\mu}^{+}$ and arrows pointing to the left $\bm{\mu}^{-}$. The diagrams
are obtained by adding arrows to the interactions in Fig.~(\protect\ref%
{fig:nrf3-partially-ordered}).}
\label{fig:nrf3-partially-ordered-rwa}
\end{figure}

Each correlation function in Eq.~(\ref{eqn:Response}) will split into $%
2^{4}=16$ terms upon substituting $\hat{\bm{\mu}}=\hat{\bm{\mu }}^{+}+\hat{%
\bm{\mu}}^{-}.$ Assuming that the system is initially in the ground state,
only two of these contributions are non-zero%
\begin{equation}
\left\langle \hat{\bm{\mu}}\hat{\bm{\mu}}\hat{\bm{\mu}}\hat{\bm{\mu}}%
\right\rangle =\left\langle \hat{\bm{\mu}}^{-}\hat{\bm{\mu}}^{+}\hat{\bm{\mu}%
}^{-}\hat{\bm{\mu}}^{+}\right\rangle +\left\langle \hat{\bm{\mu}}^{-}\hat{%
\bm{\mu}}^{-}\hat{\bm{\mu}}^{\bm{+}}\hat{\bm{\mu}}^{\bm{+}}\right\rangle .
\label{eqn:correlation function}
\end{equation}%
Substitution of Eq.~(\ref{eqn:correlation function}) into Eq.~(\ref%
{eqn:Response}) gives 
\begin{align}
\bm{S}^{(SOS)}(\tau _{4},\tau _{3},\tau _{2},\tau _{1})& =i^{3}\left[ \theta
(\tau _{43})\theta (\tau _{32})\theta (\tau _{21})\left\langle \hat{\bm{\mu}}%
^{-}(\tau _{4})\hat{\bm{\mu}}^{-}(\tau _{3})\hat{\bm{\mu}}^{+}(\tau _{2})%
\hat{\bm{\mu}}^{+}(\tau _{1})\right\rangle \right.  & & \text{(a1)}
\label{eqn:ResponseA} \\
& +\theta (\tau _{43})\theta (\tau _{32})\theta (\tau _{21})\left\langle 
\hat{\bm{\mu}}^{-}(\tau _{4})\hat{\bm{\mu}}^{+}(\tau _{3})\hat{\bm{\mu}}%
^{-}(\tau _{2})\hat{\bm{\mu}}^{+}(\tau _{1})\right\rangle  & & \text{(a2)} 
\notag \\
& -\theta (\tau _{43})\theta (\tau _{42})\theta (\tau _{21})\left\langle 
\hat{\bm{\mu}}^{-}(\tau _{3})\hat{\bm{\mu}}^{-}(\tau _{4})\hat{\bm{\mu}}%
^{+}(\tau _{2})\hat{\bm{\mu}}^{+}(\tau _{1})\right\rangle  & & \text{(b)} 
\notag \\
& \left. +\theta (\tau _{42})\theta (\tau _{23})\theta (\tau
_{41})\left\langle \hat{\bm{\mu}}^{-}(\tau _{3})\hat{\bm{\mu}}^{+}(\tau _{2})%
\hat{\bm{\mu}}^{-}(\tau _{4})\hat{\bm{\mu}}^{+}(\tau _{1})\right\rangle %
\right]  & & \text{(c)}  \notag \\
& +c.c. & &  \notag
\end{align}%

\begin{figure}[ptb]
\begin{center}
\includegraphics[clip=true, viewport=0.0in 0in 8.0in 10.0in,scale=0.5,
angle=-90]{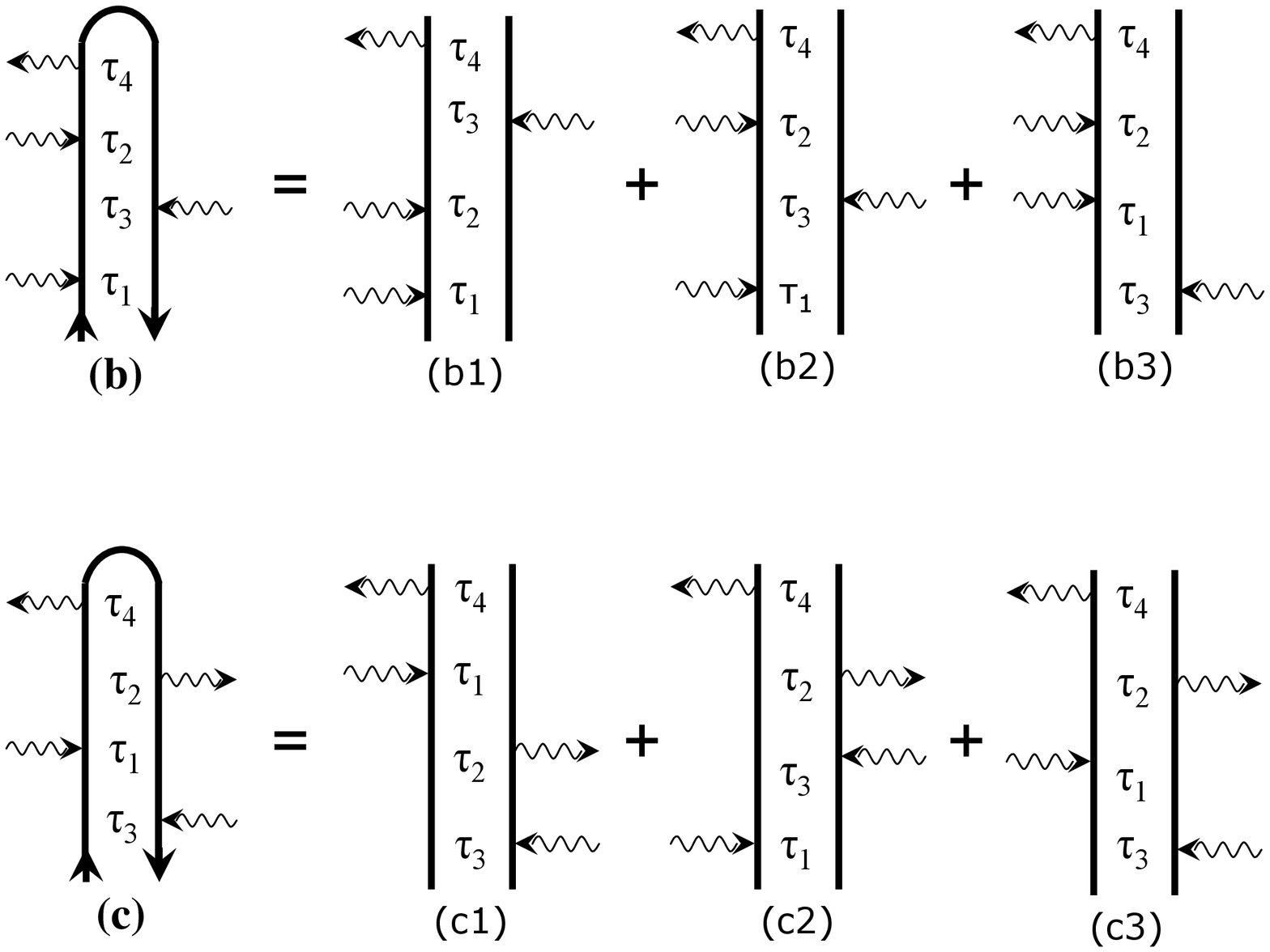}
\end{center}
\caption{Diagrams representing the fully time-ordered terms contributing to
the third-order polarization within the rotating wave approximation (Eq.~(%
\protect\ref{total S^3})). $\protect\tau_{j}$ repesent the interaction times
with the various fields. Arrows pointing to the right (left) represent $
\bm{\mu
}^{+}$ ($\bm{\mu}^{-}$). Time variables in loop diagrams \textbf{(b)%
} and \textbf{(c)} on the left are ordered in the loop. The other open
diagrams are fully ordered in physical time.}
\label{fig:nrf3-fully-ordered-rwa}
\end{figure}

The four terms represented by the diagrams in Fig.~\ref%
{fig:nrf3-partially-ordered-rwa} were obtained by taking $\hat{\bm{\mu}}%
(\tau _{4})=\hat{\bm{\mu}}^{-}(\tau _{4})$ for the last interaction,
$\hat{\bm{\mu}}(\tau _{4})=\hat{\bm{\mu}}^{+}(\tau _{4})$ gives the complex
conjugates. 
Hereafter left/right direction of the arrows corresponds to 
$\hat{\bm{\mu}}^{-}/\hat{\bm{\mu}}^{+}$ in Eq.(\ref{eqn:ResponseA}).
 Note that time-reversal symmetry implies $\left\langle \hat{%
\bm{\mu}}^{-}(\tau _{4})\hat{\bm{\mu}}^{-}(\tau _{3})\hat{\bm{\mu}}^{-}(\tau
_{2})\hat{\bm{\mu}}^{-}(\tau _{1})\right\rangle ^{\ast }=\left\langle \hat{%
\bm{\mu}}^{+}(\tau _{1})\hat{\bm{\mu}}^{+}(\tau _{2})\hat{\bm{\mu}}^{+}(\tau
_{3})\hat{\bm{\mu}}^{+}(\tau _{4})\right\rangle $. If the pulse envelopes
are much shorter than their delays, the system is forced to interact
sequentially first with pulse $\bm{k}_{1}$, then $\bm{k}_{2}$ and finally 
$\bm{k}_{3}$. This means that in the integral of Eq.~(\ref{eqn:Polarization})
one must replace $%
E(\bm{r},\tau _{j})$ with one of the $\mathcal{E}_{j}$, depending on the
time-ordering of the integration variables in real (physical) time. We note
that the first and the second terms in Eq.~(\ref{eqn:ResponseA}) impose a
full time ordering of the integration variables while the third and the
fourth terms do not. Term (b) is only partially time ordered. Depending on
the position of $\tau _{3}$ relative to the $\tau _{1}<\tau _{2}<\tau _{4}$ sequence,
the diagram can be separated into three fully time ordered terms: $\tau
_{3}<\tau _{1}$, $\tau _{1}<\tau _{3}<\tau _{2}$ or $\tau _{2}<\tau
_{3}<\tau _{4}$. Formally we do that by separating the product of step
functions\ as follows:%
\begin{equation*}
\theta (\tau _{43})\theta (\tau _{42})\theta (\tau _{21})=\theta (\tau
_{42})\theta (\tau _{21})\theta (\tau _{13})+\theta (\tau _{42})\theta (\tau
_{23})\theta (\tau _{31})+\theta (\tau _{43})\theta (\tau _{32})\theta (\tau
_{21}).
\end{equation*}%
Using this relation, diagram (b) of Fig.~\ref{fig:nrf3-partially-ordered-rwa}
is split into (b3), (b2) and (b1) as shown in the first line of Fig.~\ref%
{fig:nrf3-fully-ordered-rwa}.
The interactions on the l.h.s. of this diagrammatic equation are ordered on
the loop. On the other hand, the arrows in the open, double-sided
diagrams on the r.h.s. are ordered in real (physical)
time. All diagrams on the r.h.s. are obtained
from {\bf (b)} by moving the arrows while preserving their order along the
loop (but not in physical time!).
 Similarly we write for term (c)%
\begin{equation*}
\theta (\tau _{23})\theta (\tau _{42})\theta (\tau _{41})=\theta (\tau
_{42})\theta (\tau _{21})\theta (\tau _{13})+\theta (\tau _{42})\theta (\tau
_{23})\theta (\tau _{31})+\theta (\tau _{41})\theta (\tau _{12})\theta (\tau
_{23})
\end{equation*}%
and the diagram is split into (c2), (c3), (c1).
$\bm{S}^{(3)}$ can now be recast in the fully time-ordered form%
\begin{align}
\bm{S}^{(SOS)}(\tau _{4},\tau _{3},\tau _{2},\tau _{1})& =i^{3}\left[ \theta
(\tau _{43})\theta (\tau _{32})\theta (\tau _{21})\big\langle\hat{\bm {\mu}}%
^{-}(\tau _{4})\hat{\bm{\mu}}^{-}(\tau _{3})\hat{\bm{\mu}}^{+}(\tau _{2})\hat{%
\bm{\mu}}^{+}(\tau _{1})\big\rangle\right.  & & (\text{a1})
\label{total S^3} \\
& +\theta (\tau _{43})\theta (\tau _{32})\theta (\tau _{21})\big\langle\hat{%
\bm{\mu}}^{-}(\tau _{4})\hat{\bm{\mu}}^{+}(\tau _{3})\hat{\bm{\mu}}^{-}(\tau
_{2})\hat{\bm{\mu}}^{+}(\tau _{1})\big\rangle & & (\text{a2})  \notag \\
& -\theta (\tau _{43})\theta (\tau _{32})\theta (\tau _{21})\big\langle\hat{%
\bm{\mu}}^{-}(\tau _{3})\hat{\bm{\mu}}^{-}(\tau _{4})\hat{\bm{\mu}}^{+}(\tau
_{2})\hat{\bm{\mu}}^{+}(\tau _{1})\big\rangle & & (\text{b1})  \notag \\
& -\theta (\tau _{42})\theta (\tau _{23})\theta (\tau _{31})\big\langle\hat{%
\bm{\mu}}^{-}(\tau _{3})\hat{\bm{\mu}}^{-}(\tau _{4})\hat{\bm{\mu}}^{+}(\tau
_{2})\hat{\bm{\mu}}^{+}(\tau _{1})\big\rangle & & (\text{b2})  \notag \\
& -\theta (\tau _{42})\theta (\tau _{21})\theta (\tau _{13})\big\langle\hat{%
\bm{\mu}}^{-}(\tau _{3})\hat{\bm{\mu}}^{-}(\tau _{4})\hat{\bm{\mu}}^{+}(\tau
_{2})\hat{\bm{\mu}}^{+}(\tau _{1})\big\rangle & & (\text{b3})  \notag \\
& +\theta (\tau _{41})\theta (\tau _{12})\theta (\tau _{23})\big\langle\hat{%
\bm{\mu}}^{-}(\tau _{3})\hat{\bm{\mu}}^{+}(\tau _{2})\hat{\bm{\mu}}^{-}(\tau
_{4})\hat{\bm{\mu}}^{+}(\tau _{1})\big\rangle & & (\text{c1})  \notag \\
& +\theta (\tau _{42})\theta (\tau _{23})\theta (\tau _{31})\big\langle\hat{%
\bm{\mu}}^{-}(\tau _{3})\hat{\bm{\mu}}^{+}(\tau _{2})\hat{\bm{\mu}}^{-}(\tau
_{4})\hat{\bm{\mu}}^{+}(\tau _{1})\big\rangle & & (\text{c2})  \notag \\
& \left. +\theta (\tau _{42})\theta (\tau _{21})\theta (\tau _{13})%
\big\langle\hat{\bm{\mu}}^{-}(\tau _{3})\hat{\bm{\mu}}^{+}(\tau _{2})\hat{%
\bm{\mu}}^{-}(\tau _{4})\hat{\bm{\mu}}^{+}(\tau _{1})\big\rangle\right]  & &
(\text{c3})  \notag \\
& +c.c. & &  \notag
\end{align}%
The labels on the right correspond to the various diagrams shown in Figs.~(%
\ref{fig:nrf3-partially-ordered-rwa}) and (\ref{fig:nrf3-fully-ordered-rwa}).

Once split into fully time-ordered contributions, it is convenient to change
the integration variables in Eq.~(\ref{eqn:Polarization}) from $%
\tau_{4},\tau_{3},\tau_{2},\tau_{1}$ that label the actual interaction times
with
the fields, to the three delays $t_{3},t_{2},t_{1}$ between successive
interactions. Note that the correlation functions are invariant to time
translation $\left\langle \hat{\bm{\mu}}(\tau_{4}-\tau)\hat{\bm{\mu}}%
(\tau_{3}-\tau)\hat{\bm{\mu}}(\tau_{2}-\tau)\hat{\bm{\mu}}(\tau
_{1}-\tau)\right\rangle =\left\langle \hat{\bm{\mu}}(\tau _{4})\hat{\bm{\mu}}%
(\tau_{3})\hat{\bm{\mu}}(\tau _{2})\hat{\bm{\mu}}(\tau_{1})\right\rangle $.
Eq.~(\ref{eqn:Polarization}) thus assumes the form 
\begin{equation}
\bm{P}(\bm{r},\tau_{4})=\iiint_0^{+\infty} dt_{3}dt_{2}dt_{1}\bm{S}%
^{(3)}(t_{3},t_{2},t_{1})\bm{E}_{3}(\bm{r},\tau_{4}-t_{3})\bm{E}_{2}(\bm{r}%
,\tau_{4}-t_{2}-t_{3})\bm{E}_{1}(\bm{r},\tau_{4}-t_{1}-t_{2}-t_{3}), 
\label{eqn:Polarization3A}
\end{equation}
where $t_{1}=\tau_{2}-\tau_{1},$
$t_{2}=\tau_{3}-\tau_{2},$ $t_{3}=\tau_{4}-\tau_{3}.$ In the impulsive
limit, where all pulses are shorter than all system's response time scales,
we can substitute Eqs.~(\ref{eqn:Field})-(\ref{eqn:FieldB}) in
Eq.~(\ref{eqn:Polarization3A}) and eliminate the time integrations. This gives
\begin{gather}
\bm{P}(\bm{r},\tau_{4})=\bm{S}^{(SOS)}(t_{3},t_{2},t_{1})\mathcal{E}%
_{3}^{\lambda_{3}}(\tau_{4}-t_{3})\mathcal{E}_{2}^{\lambda_{2}}(%
\tau_{4}-t_{3}-t_{2})\mathcal{E}_{1}^{\lambda_{1}}(\tau
_{4}-t_{3}-t_{2}-t_{1})\times  \label{eqn:Polarization3B} \\
e^{i(\lambda_{1}\bm{k}_{1}+\lambda_{2}\bm{k}_{2}+\lambda _{3}\bm{k}_{3})%
\bm{r}}e^{-i(\lambda_{1}\omega_{1}+\lambda
_{2}\omega_{2}+\lambda_{3}\omega_{3})\tau_{4}}e^{i(\lambda_{1}\omega
_{1}+\lambda_{2}\omega_{2}+\lambda_{3}\omega_{3})t_{3}}e^{i(\lambda_{1}%
\omega_{1}+\lambda_{2}\omega_{2})t_{2}}e^{i\lambda_{1}\omega_{1}t_{1}} 
\notag
\end{gather}
The polarization is created along 8 possible directions $\bm{k}_{s}=\lambda_{1}%
\bm{k}_{1}+\lambda_{2}\bm{k}_{2}+\lambda _{3}\bm{k}_{3}$ with $%
\lambda_{i}=\pm1$%
\begin{equation}
\bm{P}\left( \bm{r},\tau_{4}\right) =\sum_{s=1}^{4}\bm{P}\left( \bm{k}%
_{s},\omega_{s}\right) e^{i\bm{k}_{s}\bm{r}-i\omega_{s}\tau_{4}}+c.c. 
\label{eqn:Polarization3C}
\end{equation}
where 
\begin{equation*}
\bm{P}\left( \bm{k}_{s},\omega_{s}\right) =\bm{S}_{s}^{(SOS)}\left(
t_{3},t_{2},t_{1}\right) \mathcal{E}_{3}^{\lambda_{3}}\mathcal{E}%
_{2}^{\lambda_{2}}\mathcal{E}_{1}^{\lambda_{1}}. 
\end{equation*}
$\bm{k}_{1}+\bm{k}_{2}+\bm{k}_{3}$ vanishes for the assumed dipole selection
rules
in our model. Since $\bm{P}\left( -\bm{k}_{s},-\omega_{s}\right) =\bm{P}%
^{\ast}\left( \bm{k}_{s},\omega_{s}\right) $, we are left with three
independent combinations $\bm{k}_{I}\equiv-\bm{k}_{1}+\bm{k}_{2}+\bm{k}_{3}$%
, $\bm{k}_{II}\equiv+\bm{k}_{1}-\bm{k}_{2}+\bm{k}_{3}$, $\bm{k}_{III}\equiv+%
\bm{k}_{1}+\bm{k}_{2}-\bm{k}_{3}$:%
\begin{equation*}
\bm{S}^{(SOS)}(t_{3},t_{2},t_{1})=\bm{S}_{I}^{(SOS)}(t_{3},t_{2},t_{1})+%
\bm{S}_{II}^{(SOS)}(t_{3},t_{2},t_{1})+\bm{S}%
_{III}^{(SOS)}(t_{3},t_{2},t_{1}). 
\end{equation*}
\begin{figure}[ptb]
\begin{center}
\includegraphics[clip=true, viewport=1.0in 2.5in 5.5in
8.0in,scale=0.65,angle=-90]{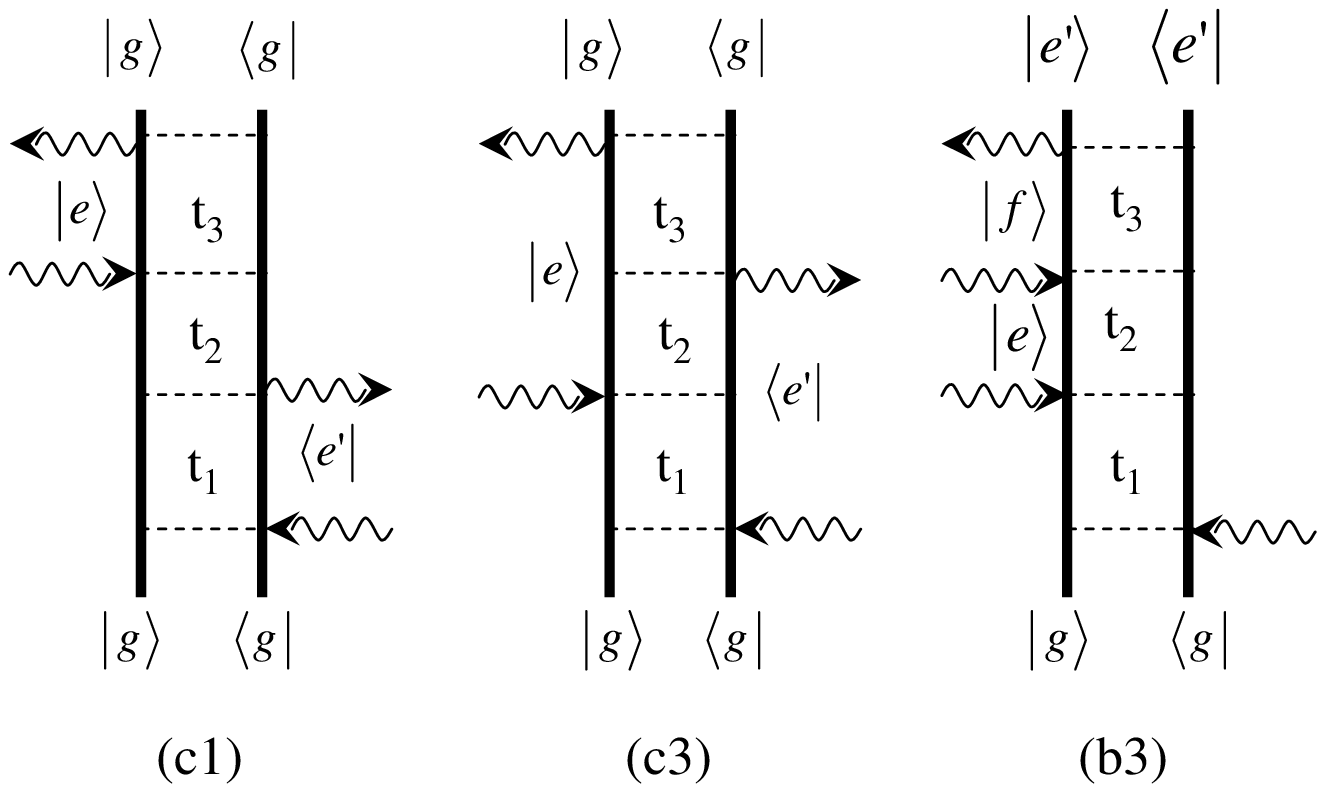}
\end{center}
\caption{Feynman diagrams for the $\bm{k}_{I}$ technique~ (Eq.~\protect\ref%
{tech I}).}
\label{fig:nrf3-k1}
\end{figure}
\begin{figure}[ptbp]
\begin{center}
\includegraphics[clip=true, viewport=1.0in 2.5in 4.5in
8.0in,scale=0.65,angle=-90]{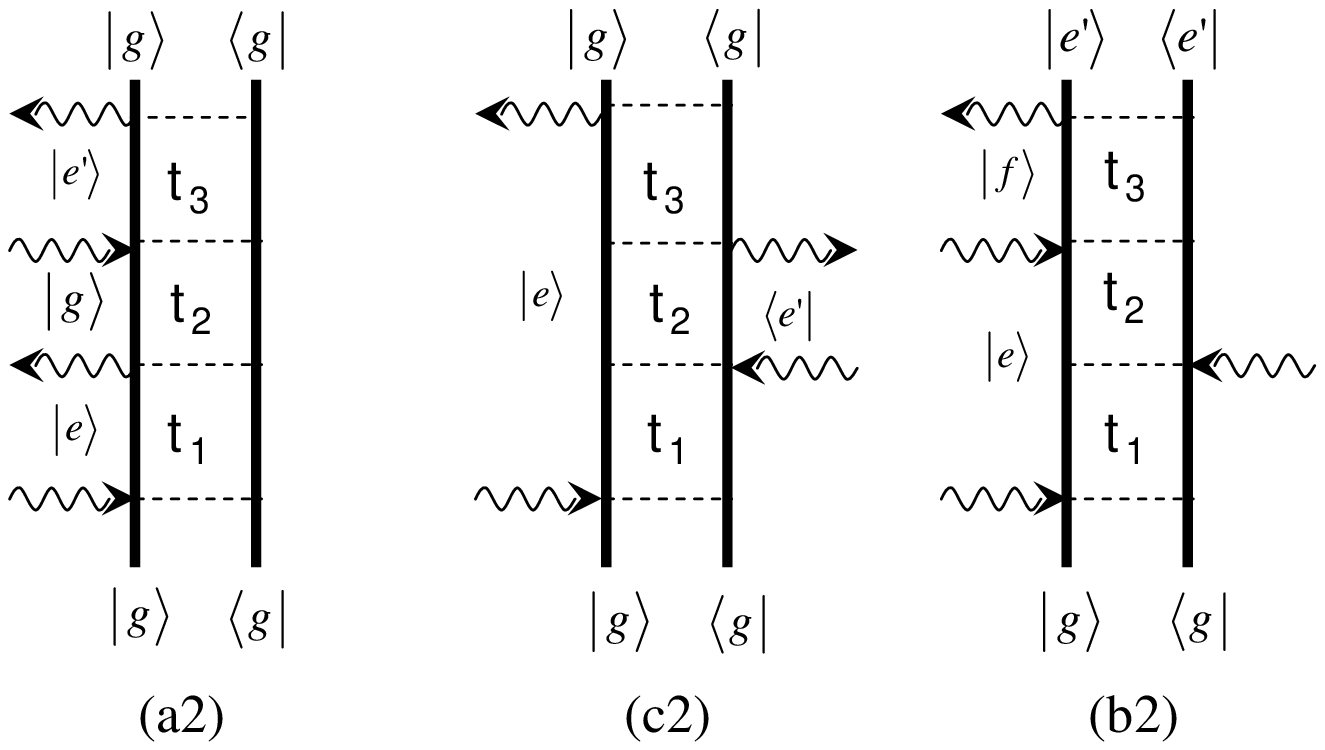}
\end{center}
\caption{Feynman diagrams for the $\bm{k}_{II}$ technique~ (Eq.~\protect\ref%
{tech II}).}
\label{fig:nrf3-k2}
\end{figure}
\begin{figure}[ptbp]
\begin{center}
\includegraphics[clip=true, viewport=1.0in 2.5in 5.5in
8.5in,scale=0.65,angle=-90]{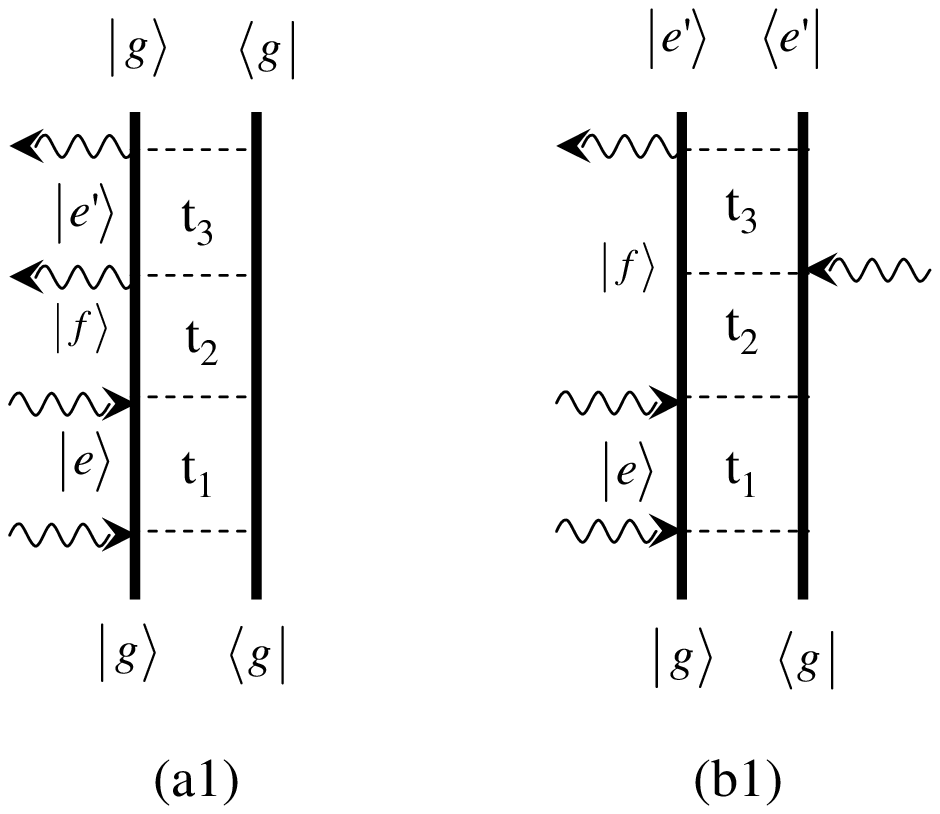}
\end{center}
\caption{Feynman diagrams for the $\bm{k}_{III}$ technique~(Eq.~\protect\ref%
{tech III}).}
\label{fig:nrf3-k3}
\end{figure}

We can classify the diagrams in Fig.~\ref{fig:nrf3-fully-ordered-rwa}
according to the directions of the arrows: arrow pointing to the right
(left) 
represents $+\bm{k}$ ($-\bm{k}$), arrows are read 
from the bottom up on either side. We obtain for 
$\bm{k}_{I}$
(Fig.~\ref{fig:nrf3-k1})%
\begin{align}
\bm{S}_{I}^{(SOS)} & =i^{3}\theta(t_{1})\theta(t_{2})\theta (t_{3})\left[
\left\langle \hat{\bm{\mu}}^{-}(0)\hat{\bm{\mu}}^{+}(t_{1}+t_{2})\hat{%
\bm{\mu}}^{-}(t_{1}+t_{2}+t_{3})\hat{\bm{\mu}}^{+}(t_{1})\right\rangle
\right. & & (\text{c3})  \label{tech I} \\
& +\left\langle \hat{\bm{\mu}}^{-}(0)\hat{\bm{\mu}}^{+}(t_{1})\hat{\bm{\mu}}%
^{-}(t_{1}+t_{2}+t_{3})\hat{\bm{\mu}}^{+}(t_{1}+t_{2})\right\rangle & & (%
\text{c1})  \notag \\
& \left. -\left\langle \hat{\bm{\mu}}^{-}(0)\hat{\bm{\mu}}%
^{-}(t_{1}+t_{2}+t_{3})\hat{\bm{\mu}}^{+}(t_{1}+t_{2})\hat{\bm{\mu}}%
^{+}(t_{1})\right\rangle \right] . & & (\text{b3})  \notag
\end{align}
For the $\bm{k}_{II}$ technique we similarly
have (Fig.~\ref{fig:nrf3-k2}):%
\begin{align}
\bm{S}_{II}^{(SOS)} & =i^{3}\theta(t_{1})\theta(t_{2})\theta (t_{3})\left[
\left\langle \hat{\bm{\mu}}^{-}(t_{1}+t_{2}+t_{3})\hat{\bm{\mu}}%
^{+}(t_{1}+t_{2})\hat{\bm{\mu}}^{-}(t_{1})\hat{\bm{\mu}}^{+}(0)\right\rangle
\right. & & (\text{a2})  \label{tech II} \\
& +\left\langle \hat{\bm{\mu}}^{-}(t_{1})\hat{\bm{\mu}}^{+}(t_{1}+t_{2})\hat{%
\bm{\mu}}^{-}(t_{1}+t_{2}+t_{3})\hat{\bm {\mu}}^{+}(0)\right\rangle & & (%
\text{c2})  \notag \\
& -\left. \left\langle \hat{\bm{\mu}}^{-}(t_{1})\hat{\bm {\mu}}%
^{-}(t_{1}+t_{2}+t_{3})\hat{\bm{\mu}}^{+}(t_{1}+t_{2})\hat{\bm{\mu}}%
^{+}(0)\right\rangle \right] & & (\text{b2})  \notag
\end{align}
Finally $\bm{k}_{III}$ is given by
 (Fig.~\ref{fig:nrf3-k3}): 
\begin{align}
\bm{S}_{III}^{(SOS)} & =i^{3}\theta(t_{1})\theta(t_{2})\theta (t_{3})\left[
\left\langle \hat{\bm{\mu}}^{-}(t_{1}+t_{2}+t_{3})\hat{\bm{\mu}}%
^{-}(t_{1}+t_{2})\hat{\bm{\mu}}^{+}(t_{1})\hat{\bm{\mu}}^{+}(0)\right\rangle
\right. & & (\text{a1})  \label{tech III} \\
& -\left. \left\langle \hat{\bm{\mu}}^{-}(t_{1}+t_{2})\hat{\bm{\mu}}%
^{-}(t_{1}+t_{2}+t_{3})\hat{\bm{\mu}}^{+}(t_{1})\hat{\bm{\mu}}%
^{+}(0)\right\rangle \right] & & (\text{b1})  \notag
\end{align}
Each term is labelled according to Eq.~(\ref{total S^3}). Eqs.~(\ref{tech I}-%
\ref{tech III}) can be used to express the third order SOS response in terms
of transition dipoles, system frequencies and dephasing rates (see App.~\ref%
{I_expressions} and Sec.~\ref{Sec:2D}). 

In the next section we employ the
EOM approach to derive the alternative QP expressions for these signals. These
will then be connected with the current SOS expressions in Section~\ref%
{Sec:Connection}.

\section{Quasiparticle expressions for Wannier excitons in semiconductors
\label{Sec:QP}}

Interband transitions in semiconductors may be described by the two-band
many-electron Hamiltonian:\cite{AxtMukamel1998rev,HaugKoch2004}%
\begin{equation}
\hat{H}_{T}=\hat{H}_{0}+\hat{H}_{C}+\hat{H}_{I}~,   \label{start H}
\end{equation}
with the single-particle part%
\begin{equation*}
\hat{H}_{0}=\sum_{m_{1},n_{1}}t_{m_{1},n_{1}}^{\left( 1\right)
}c_{m_{1}}^{\dagger}c_{n_{1}}+\sum_{m_{2},n_{2}}t_{m_{2},n_{2}}^{\left(
2\right) }d_{m_{2}}^{\dagger}d_{n_{2}}~, 
\end{equation*}
where $c^{\dagger}$ create electrons and $d^{\dagger}$ create holes. The
Coulomb interaction is:%
\begin{align*}
\hat{H}_{C} & =\frac{1}{2}%
\sum_{m_{1},n_{1},k_{1},l_{1}}V_{m_{1}n_{1}k_{1}l_{1}}^{\left( 1\right)
}c_{m_{1}}^{\dagger}c_{n_{1}}^{\dagger}c_{k_{1}}c_{l_{1}}+\frac{1}{2}%
\sum_{m_{2},n_{2},k_{2},l_{2}}V_{m_{2}n_{2}k_{2}l_{2}}^{\left( 2\right)
}d_{m_{2}}^{\dagger}d_{n_{2}}^{\dagger}d_{k_{2}}d_{l_{2}} \\
& -%
\sum_{m_{1},n_{2},k_{2},l_{1}}W_{m_{1}n_{2}l_{1}k_{2}}c_{m_{1}}^{%
\dagger}d_{n_{2}}^{\dagger}d_{k_{2}}c_{l_{1}}~,
\end{align*}
while 
\begin{equation*}
\hat{H}_{I}=- \sum_{m_{1},m_{2}}\left( \bm{E}^{+}\left( t\right) \bm{\mu}%
_{m_{1}m_{2}}^{\ast}c_{m_{1}}^{\dagger}d_{m_{2}}^{\dagger }+\bm{E}^{-}\left(
t\right) \bm{\mu}_{m_{1}m_{2}}d_{m_{2}}c_{m_{1}}\right) , 
\end{equation*}
is the dipole interaction with light, and the optical electric field $E$ will be
treated as a scalar for simplicity. $\hat{H}_{T}$ can describe both bulk and
low-dimensional semiconductor systems. All the steps in this Section are
independent of the single-electron basis used. $\hat{H}_{0}$ would be diagonal in
the basis of the system's single-particle eigenstates, i.e., $%
t_{m1,n1}^{\left( i\right) }=\varepsilon_{m_{1}}^{\left( i\right) }\delta
_{m_{1}n_{1}}$. In this paper we focus on the coherent response and
we neglect coupling with phonons,
which would result in additional, relevant dynamical variables and new
contributions
to the response function. \cite{ChernyakZhangMukamel98,Mukamel_inzyss1993}
The SOS and QP pictures should be equivalent also when dephasing is
included. In that case, however, the theory becomes more complicated. For the
sake of simplicity and transparency we restrict the following analysis to the
coherent response, where we do not include phonons explicitly.
Dephasing effects, necessary for a realistic description, will be simply 
introduced by adding imaginary parts to excitonic frequencies.

To introduce the exciton representation we define electron-hole operators:%
\cite{ChernyakMukamel1996}%
\begin{equation*}
\hat{B}_{m}\equiv d_{m_{2}}c_{m_{1}},~~\hat{B}_{m}^{\dagger}\equiv
c_{m_{1}}^{\dagger}d_{m_{2}}^{\dagger},
\end{equation*}
where we have employed shorthand notation for pairs of indices: $m\equiv
\left( m_{1},m_{2}\right) $. Using these operators we construct an effective
Hamiltonian $\hat{H}$ (see App.~\ref{Hamiltonian}):%
\begin{equation}
\hat{H}=\sum_{mn}h_{mn}\hat{B}_{m}^{\dagger}\hat{B}_{n}+\sum_{mnkl}U_{mnkl}%
\hat{B}_{m}^{\dagger}\hat{B}_{n}^{\dagger}\hat{B}_{k}\hat{B}%
_{l}-\sum_{m}\left( \bm{E}^{+}\left( t\right) \bm{\mu}_{m}^{\ast}\hat{B}%
_{m}^{\dagger}+\bm{E}^{-}\left( t\right) \bm{\mu}_{m}\hat{B}_{m}\right) ~. 
\label{main H}
\end{equation}
The Hamiltonians $\hat{H}$ and $\hat{H}_{T}$ are equivalent in the single and
double
excitations subspace, which is relevant for the response to third order in $E
$.\cite{mukbook} This transformation from fermion to exciton variables is
crucial for our approach, since it allows us to view the electronic degrees
of freedom as a system of coupled oscillators. The parameters of the
transformed Hamiltonian $\hat{H}$ are given by:%
\begin{align}
h_{mn} & =t_{m_{1},n_{1}}^{\left( 1\right)
}\delta_{m_{2}n_{2}}+t_{m_{2},n_{2}}^{\left( 2\right) }\delta_{m_{1}n_{1}}%
          -W_{m_{1}m_{2}n_{1}n_{2}},  \label{int defin} \\
U_{mnkl} & =-\frac{1}{4}\left[ t_{m_{1},k_{1}}^{\left( 1\right)
}\delta_{m_{2}k_{2}}\delta_{n_{1}l_{1}}\delta_{n_{2}l_{2}}+t_{m_{2},k_{2}}^{%
\left( 2\right)
}\delta_{m_{1}k_{1}}\delta_{n_{1}l_{1}}\delta_{n_{2}l_{2}}+\right.  \notag \\
& \left. t_{n_{1},l_{1}}^{\left( 1\right)
}\delta_{m_{1}k_{1}}\delta_{m_{2}k_{2}}\delta_{n_{2}l_{2}}+t_{n_{2},l_{2}}^{%
\left( 2\right) }\delta_{m_{1}k_{1}}\delta_{m_{2}k_{2}}\delta_{n_{1}l_{1}}
\right] +  \notag \\
& \frac{1}{4}\left[ V_{m_{1}n_{1}k_{1}l_{1}}^{(1)}\delta_{m_{2}k_{2}}%
\delta_{n_{2}l_{2}}+V_{m_{2}n_{2}k_{2}l_{2}}^{(2)}\delta_{m_{1}k_{1}}%
\delta_{n_{1}l_{1}}\right] .  \notag
\end{align}
The commutation relations for the $\hat{B}$ operators can be obtained using
the elementary fermion anticommutators: $\left[ c_{m_{1}}^{\dagger},c_{k_{1}}%
\right] _{+}=\delta_{m_{1},k_{1}}$.\ Within the subspace of $\left\vert
0\right\rangle $ and $\hat{B}_{i}^{\dagger}\left\vert 0\right\rangle $
states (i.e., the ground state and single excitations), we get \cite%
{ChernyakMukamel1996}%
\begin{equation}
\left[ \hat{B}_{m},\hat{B}_{n}^{\dagger}\right] =\delta_{mn}-2\sum _{pq}%
\mathcal{P}_{mnpq}\hat{B}_{p}^{\dagger}\hat{B}_{q}~,   \label{comm rel}
\end{equation}
where $\delta_{mn}=\delta_{m_{1}n_{1}}\delta_{m_{2}n_{2}}$ and%
\begin{equation}
\mathcal{P}_{mnpq}=\frac{1}{2}\delta_{m_{1}q_{1}}\delta_{p_{1}n_{1}}\delta_{m_{2}p_{2}}%
\delta_{n_{2}q_{2}}+\frac{1}{2}\delta_{m_{2}q_{2}}\delta_{p_{2}n_{2}}%
\delta_{m_{1}p_{1}}\delta_{n_{1}q_{1}}\text{.}   \label{P def}
\end{equation}
Eqs.~(\ref{comm rel}) and (\ref{P def}) are obtained in a similar way to (%
\ref{main H}) and (\ref{int defin}). \ Terms with additional $\hat{B}%
_{i}^{\dagger}\hat{B}_{j}$ pairs (e.g. $\hat{B}^{\dagger}\hat{B}^{\dagger}%
\hat{B}\hat{B}$) are neglected in (\ref{comm rel}), because they would
introduce corrections higher than $O\left( E^{3}\right) $ to the nonlinear
response. Note the symmetry $\mathcal{P}_{mnpq}=\mathcal{P}_{mnqp}$.

Using Eqs.~(\ref{main H}) and (\ref{comm rel}) we obtain the nonlinear
exciton equations (see Appendix~\ref{EOM}) for single-exciton variables $%
\left\langle \hat{B}_{m}\right\rangle $:\cite%
{spanomuk91a,Mukamel_inzyss1993,ChernYoko1998,mukamel_annrev2000}%
\begin{align}
i\frac{d\left\langle \hat{B}_{m}\right\rangle }{dt} &
=\sum_{n}h_{mn}\left\langle \hat{B}_{n}\right\rangle -\bm{\mu}_{m}^{\ast }%
\bm{E}^{+}\left( t\right) +\sum_{nkl}V_{mnkl}\left\langle \hat {B}%
_{n}\right\rangle ^{\ast}\left\langle \hat{B}_{k}\hat{B}_{l}\right\rangle
\label{EOM 1} \\
& +2\bm{E}^{+}\left( t\right) \sum_{npq}\mathcal{P}_{mnpq}\left\langle \hat{B%
}_{n}\right\rangle ^{\ast}\left\langle \hat{B}_{q}\right\rangle \bm{\mu}%
_{p}^{\ast},  \notag
\end{align}
where $V$ is given by 
\begin{equation}
V_{nmpq}=2U_{nmpq}-2\sum_{l}\mathcal{P}_{nmlp}h_{lq}-2\sum_{k,l}\mathcal{P}%
_{nmkl}U_{klpq}.   \label{eqn:anh def}
\end{equation}
Here $Y_{mn}\equiv\left\langle \hat{B}_{m}\hat{B}_{n}\right\rangle $ are
two-exciton variables. The Heisenberg equations give:%
\begin{align}
i\frac{dY_{mn}}{dt} & =\sum_{kl}h_{mn,kl}^{\left( Y\right) }Y_{kl}-\bm{E}%
^{+}\left( t\right) \left( \left\langle \hat{B}_{n}\right\rangle \bm{\mu}%
_{m}^{\ast}+\left\langle \hat{B}_{m}\right\rangle \bm{\mu}_{n}^{\ast}\right)
\label{EOM 2} \\
& +2\bm{E}^{+}\left( t\right) \sum_{k,l}\mathcal{P}_{mnkl}\left\langle \hat{B%
}_{k}\right\rangle \bm{\mu}_{l}^{\ast},  \notag
\end{align}

Calculating the optical response by numerical integration of these equations%
\cite{SiehMeier1999,Lijun2006} is straightforward but numerically expensive.
An alternative, more tractable approach, which further provides a better
insight into the nature of the response, is\ to integrate the equations
formally using one-exciton Green's functions $G\left( t\right) $ and exciton
scattering matrix $\Gamma\left( t\right) $. The scattering matrix depends on
quasiparticle statistics through the $\mathcal{P}$ matrix (Eqs.~\ref{Gamma
def},~\ref{Gamma def temp}) as well as on exciton-exciton coupling. This
results in closed quasiparticle expressions for the 3rd order contributions $%
S_{I}$, $S_{II}$ and $S_{III}$ to the response function (for details see
Appendix \ref{Resp fun} and Ref.~\onlinecite{AbramaviciusMukamel_time2005})%
\begin{align}
& \bm{S}_{^{I}}^{\left( QP\right) }(\tau_{4},\tau_{3},\tau_{2},\tau_{1})=
\label{S t 1} \\
& -2\theta\left( \tau_{43}\right) \theta\left( \tau_{32}\right) \theta\left(
\tau_{21}\right) \sum_{n_{4}...n_{1}}\bm{\mu}_{n_{4}}\bm{\mu}_{n_{3}}^{\ast}%
\bm{\mu}_{n_{2}}^{\ast}\bm{\mu }_{n_{1}}\int\limits_{-\infty}^{\tau_{43}}d%
\tau_{s}^{\prime\prime}\int\limits_{0}^{\tau_{s}^{\prime\prime}}d\tau_{s}^{%
\prime}\sum_{n_{1}^{\prime},n_{2}^{\prime},n_{3}^{\prime},n_{4}^{\prime}}%
\times  \notag \\
& \Gamma_{n_{4}^{\prime}n_{1}^{\prime}n_{3}^{\prime}n_{2}^{\prime}}\left(
\tau_{s}^{\prime\prime}-\tau_{s}^{\prime}\right)
G_{n_{4}n_{4}^{\prime}}\left( \tau_{s}^{\prime}\right)
G_{n_{3}^{\prime}n_{3}}\left( \tau _{43}-\tau_{s}^{\prime\prime}\right)
G_{n_{2}^{\prime}n_{2}}\left( \tau _{42}-\tau_{s}^{\prime\prime}\right)
G_{n_{1}^{\prime}n_{1}}^{\ast}\left( \tau_{41}-\tau_{s}^{\prime}\right) , 
\notag
\end{align}
where $\tau_{43}=\tau_{4}-\tau_{3},$ etc. and $G_{mn}\left( t\right)
=-i\theta\left( t\right) \left[ \exp\left( -iht\right) \right] _{mn}$.

The response functions for the other phase-matching directions can be
derived along the same lines. We get%
\begin{align}
& \bm{S}_{II}^{\left( QP\right) }(\tau_{4},\tau_{3},\tau_{2},\tau_{1})=
\label{S t 2} \\
& -2\theta\left( \tau_{43}\right) \theta\left( \tau_{32}\right) \theta\left(
\tau_{21}\right) \sum_{n_{4}...n_{1}}\bm{\mu}_{n_{4}}\bm{\mu}_{n_{3}}^{\ast}%
\bm{\mu}_{n_{2}}\bm{\mu }_{n_{1}}^{\ast}\int\limits_{-\infty}^{\tau_{43}}d%
\tau_{s}^{\prime\prime}\int\limits_{0}^{\tau_{s}^{\prime\prime}}d\tau_{s}^{%
\prime}\sum_{n_{4}^{\prime}...n_{1}^{\prime}}\times  \notag \\
& \Gamma_{n_{4}^{\prime}n_{2}^{\prime}n_{3}^{\prime}n_{1}^{\prime}}(\tau
_{s}^{\prime\prime}-\tau_{s}^{\prime})G_{n_{4}n_{4}^{\prime}}(\tau_{s}^{%
\prime})G_{n_{3}^{\prime}n_{3}}(\tau_{43}-\tau_{s}^{\prime\prime})G_{n_{2}^{%
\prime}n_{2}}^{\ast}(\tau_{42}-\tau_{s}^{\prime})G_{n_{1}^{\prime
}n_{1}}(\tau_{41}-\tau_{s}^{\prime\prime}),  \notag
\end{align}
and:%
\begin{align}
& \bm{S}_{III}^{\left( QP\right) }(\tau_{4},\tau_{3},\tau_{2},\tau_{1})=
\label{S t 3} \\
& -2\theta\left( \tau_{43}\right) \theta\left( \tau_{32}\right) \theta\left(
\tau_{21}\right) \sum_{n_{4}...n_{1}}\bm{\mu}_{n_{4}}\bm{\mu}_{n_{3}}\bm{\mu}%
_{n_{2}}^{\ast}\bm{\mu }_{n_{1}}^{\ast}\int\limits_{-\infty}^{\tau_{42}}d%
\tau_{s}^{\prime\prime}\int\limits_{0}^{\tau_{s}^{\prime\prime}}d\tau_{s}^{%
\prime}\sum_{n_{4}^{\prime}...n_{1}^{\prime}}\times  \notag \\
& \Gamma_{n_{4}^{\prime}n_{3}^{\prime}n_{2}^{\prime}n_{1}^{\prime}}(\tau
_{s}^{\prime\prime}-\tau_{s}^{\prime})G_{n_{4}n_{4}^{\prime}}(\tau_{s}^{%
\prime})G_{n_{3}^{\prime}n_{3}}^{\ast}(\tau_{43}-\tau_{s}^{%
\prime})G_{n_{2}^{\prime}n_{2}}(\tau_{42}-\tau_{s}^{\prime\prime})G_{n_{1}^{%
\prime }n_{1}}(\tau_{41}-\tau_{s}^{\prime\prime})\text{.}  \notag
\end{align}
Just as in the SOS case, time translation symmetry implies that these response 
functions only depend on the three
pulse delays $t_{3},t_{2},t_{1}$. Eqs.~(\ref{S t 1}-\ref{S t 3}) will be
used next to connect the QP and the SOS pictures.

\section{Connecting the sum-over-states and the quasiparticle pictures\label%
{Sec:Connection}}

We first recast Eqs.~(\ref{tech I}-\ref{tech III}) using Green's functions
(in all expressions $t_{1}>0$, $t_{2}>0$, $t_{3}>0$):%
\begin{align}
\bm{S}_{I}^{(SOS)} = & -\left\langle \hat{\bm{\mu}}^{-}\hat {G}%
^{\dagger}(t_{1}+t_{2}+t_{3})\hat{\bm{\mu}}^{-}\mathcal{\hat{G}}(t_{3})\hat{%
\bm{\mu}}^{+}\hat{G}(t_{2})\hat{\bm{\mu}}^{+}\right\rangle & & \text{(b3)}
\label{S 1 subst} \\
& -\left\langle \hat{\bm{\mu}}^{-}\hat{G}^{\dagger}(t_{1}+t_{2})\hat{\bm{\mu}%
}^{+}\mathbb{\hat{G}}^{\mathbb{\dagger}}(t_{3})\hat{\bm{\mu}}^{-}\hat{G}%
(t_{2}+t_{3})\hat{\bm{\mu}}^{+}\right\rangle & & \text{(c3)}  \notag \\
& -\left\langle \hat{\bm{\mu}}^{-}\hat{G}^{\dagger}\left( t_{1}\right) \hat{%
\bm{\mu}}^{+}\mathbb{\hat{G}}^{\mathbb{\dagger}}(t_{2}+t_{3})\hat{\bm{\mu}}%
^{-}\hat{G}(t_{3})\hat{\bm{\mu}}^{+}\right\rangle , & & \text{(c1)}  \notag
\\
\bm{S}_{II}^{(SOS)} = & -\left\langle \hat{\bm{\mu}}^{-}\hat{G}(t_{3})\hat{%
\bm{\mu}}^{+}\mathbb{\hat{G}}(t_{2})\hat{\bm{\mu}}^{-}\hat{G}(t_{1})\hat{%
\bm{\mu}}^{+}\right\rangle & & \text{(a2)}  \label{S 2 subst} \\
& -\left\langle \hat{\bm{\mu}}^{-}\hat{G}^{\dagger}(t_{2}+t_{3})\hat{\bm{\mu}%
}^{-}\mathcal{\hat{G}}(t_{3})\hat{\bm{\mu}}^{+}\hat{G}(t_{1}+t_{2})\hat{%
\bm{\mu}}^{+}\right\rangle & & \text{(b2)}  \notag \\
& -\left\langle \hat{\bm{\mu}}^{-}\hat{G}^{\dagger}(t_{2})\hat{\bm{\mu}}^{+}%
\mathbb{\hat{G}}^{\mathbb{\dagger}}(t_{3})\hat{\bm{\mu}}^{-}\hat{G}%
(t_{1}+t_{2}+t_{3})\hat{\bm{\mu}}^{+}\right\rangle , & & \text{(c2)}  \notag
\\
\bm{S}_{III}^{(SOS)} = & -\left\langle \hat{\bm{\mu}}^{-}\hat{G}(t_{3})\hat{%
\bm{\mu}}^{-}\mathcal{\hat{G}}(t_{2})\hat{\bm{\mu}}^{+}\hat{G}(t_{1})\hat{%
\bm{\mu}}^{+}\right\rangle & & \text{(a1)}  \label{S 3 subst} \\
& -\left\langle \hat{\bm{\mu}}^{-}\hat{G}^{\dagger}(t_{3})\hat{\bm{\mu}}^{-}%
\mathcal{\hat{G}}(t_{2}+t_{3})\hat{\bm{\mu}}^{+}\hat{G}(t_{1})\hat{\bm{\mu}}%
^{+}\right\rangle , & & \text{(b1)}  \notag
\end{align}
Here $\hat{G}(t)\equiv-i\theta(t)\exp(-i\hat{H}t)$ and $\hat{G}^{\dag
}(t)\equiv+i\theta(t)\exp(i\hat{H}t)$ represent the retarded and the
advanced Green's function respectively; $\mathbb{\hat{G}}$, $\hat{G}$ and $%
\mathcal{\hat{G}}$ describe the evolution within the ground-state,
single-exciton and double-exciton blocks of the Hamiltonian (Eq.~\ref{main H}%
) respectively. We also set the ground state energy $\varepsilon_{g}$ to
zero.

Our goal is to show the equivalence of the QP and SOS pictures by deriving
Eqs.~(\ref{S t 1}-\ref{S t 3}) from Eqs.~(\ref{S 1 subst}-\ref{S 3 subst}).
To that end we adopt a harmonic reference system of noninteracting quasiparticle and
expand the SOS response in anharmonicities. Harmonic oscillators are linear,
and their nonlinear response vanishes identically \cite%
{spanomuk89,KuhnChernyak96,Mukamel_inzyss1993,mukbook}, as can be easily
seen from the Heisenberg equations of motion.\ This means that the various
Liouville space pathways for all nonlinear response function interfere
destructively. Exploiting this property in the following derivation,
we show that the quasiparticle physical picture has built-in cancellations
in the reference harmonic system.

We shall use the Dyson equation for the two particle Green's function, also
known as the Bethe-Salpeter equation\cite{leegwaterMukamel92}%
\begin{equation}
\mathcal{\hat{G}}\left( \omega\right) =\mathcal{\hat{G}}_{0}\left(
\omega\right) +\mathcal{\hat{G}}_{0}\left( \omega\right) \Gamma\left(
\omega\right) \mathcal{\hat{G}}_{0}\left( \omega\right) ~, 
\label{eqn:Bethe}
\end{equation}
or in the time domain:%
\begin{equation}
\mathcal{\hat{G}}(\tau)=\mathcal{\hat{G}}_{0}(\tau)+\int_{0}^{\tau}d\tau^{%
\prime}\int_{0}^{\tau^{\prime}}d\tau^{\prime\prime}\mathcal{\hat{G}}%
_{0}\left( \tau-\tau^{\prime}\right) \Gamma\left(
\tau^{\prime}-\tau^{\prime\prime}\right) \mathcal{\hat{G}}%
_{0}(\tau^{\prime\prime}).   \label{B S time}
\end{equation}
$\mathcal{\hat{G}}_{0}$ is taken to be the Green's function of a doubly
excited, harmonic system. It can be factorized into the product of a
single-exciton Green's functions 
\begin{equation*}
\mathcal{\hat{G}}_{0}(\tau)_{n_{1}n_{2}n_{3}n_{4}}=i\hat{G}%
_{n_{1}n_{3}}(\tau)\hat{G}_{n_{2}n_{4}}(\tau). 
\end{equation*}
The exciton scattering matrix $\Gamma$ is defined by Eq.~(\ref{B S time}).

\begin{figure}[ptb]
\begin{center}
\includegraphics[clip=true, viewport=0.0in 3.0in 6.0in
10.0in,scale=0.6,angle=-90]{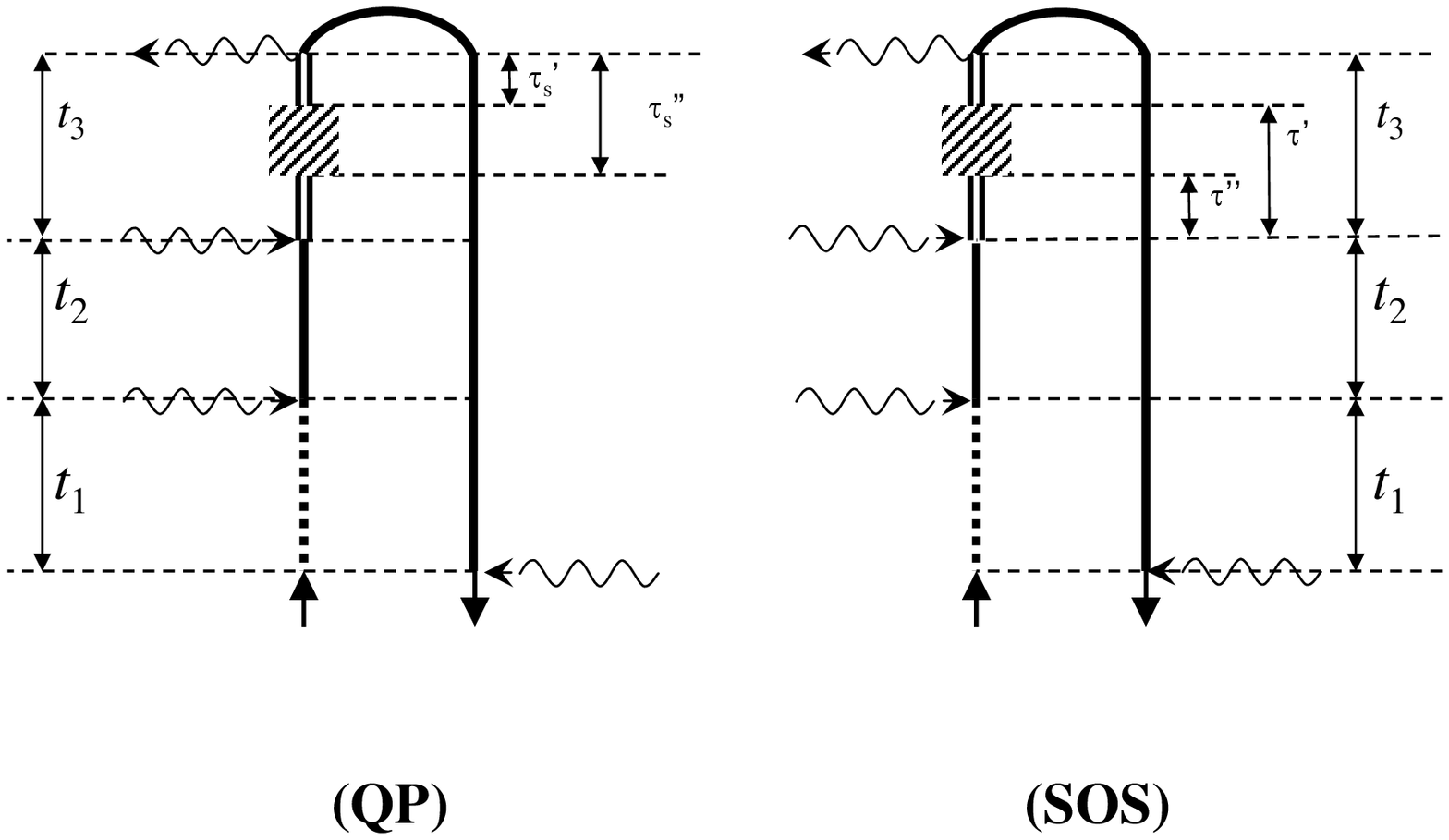}
\end{center}
\caption{Loop diagrams showing the equivalence of the $S_{I}$ expressions in
the QP (Eq.~\protect\ref{S t 1}) and SOS (Eq.~\protect\ref{S 1 subst 2})
pictures. Dotted, single and double lines show ground, single exciton and
double exciton states' evolution respectively. Dashed region represents
scattering matrix.}
\label{Darius_K1}
\end{figure}

Let us start with the $\bm{S}_{I}$ technique and show the equivalence of
Eq.~(\ref{S 1 subst}) to (\ref{S t 1}). The second and third terms of Eq.~(%
\ref{S 1 subst}) (diagrams (c1) and (c3) in Fig.~\ref{fig:nrf3-k1}) are
purely harmonic, independent on the quasiparticle interactions. This is a
direct result of the ordering of $\bm{\mu}^{\pm}$, whereby the system only
evolves in the ground and first excited state. Exciton-exciton interactions
influence the evolution only in the second excited manifold. The first term
in Eq.~(\ref{B S time}), i.e. $\mathcal{\hat{G}}_{0}$, represents harmonic
evolution in the two-exciton manifold. Thus the first term of Eq.~(\ref{S 1
subst}) with $\mathcal{\hat{G}}$ replaced by $\mathcal{\hat{G}}_{0}$ must
cancel the other two terms, because the nonlinear response of a harmonic
system vanishes. Substituting the second term from Eq.~(\ref{B S time}) in
Eq.~(\ref{S 1 subst}) we obtain a single term for $\bm{S}_{I}$:%
\begin{align}
\bm{S}_{I}^{\left( SOS\right) } & =-\theta\left( t_{3}\right) \theta\left(
t_{2}\right) \theta\left( t_{1}\right)
\int_{0}^{t_{3}}d\tau^{\prime}\int_{0}^{\tau^{\prime}}d\tau^{\prime\prime}%
\times  \label{S 1 subst 2} \\
& \left\langle \hat{\bm{\mu}}^{-}\hat{G}^{\dagger}(t_{1}+t_{2}+t_{3})\hat{%
\bm{\mu}}^{-}\mathcal{\hat{G}}_{0}(t_{3}-\tau^{\prime
})\Gamma(\tau^{\prime}-\tau^{\prime\prime})\mathcal{\hat{G}}_{0}(\tau
^{\prime\prime})\hat{\bm{\mu}}^{+}\hat{G}(t_{2})\hat{\bm{\mu}}%
^{+}\right\rangle ,  \notag
\end{align}
The equivalence of the Eqs.~(\ref{S 1 subst 2}) and (\ref{S t 1}) can
be directly seen using the diagrams shown in Fig.~(\ref{Darius_K1}). In
these diagrams the scattering matrix $\Gamma$ is represented by dashed
regions. Note that $G\left( t\right) G^{\dagger}\left( t\right)
=\theta\left( t\right) \exp(-iht)\exp(iht)=\theta\left( t\right) $. The QP
diagram in Fig.~(\ref{Darius_K1}) is obtained from the SOS one by changing
the integration variables $\tau^{\prime}=t_{3}-\tau_{s}^{\prime}$ and $%
\tau^{\prime\prime}=t_{3}-\tau_{s}^{\prime\prime}.$ This completes the
derivation of the QP expression for $S_{I}$ (Eq.~\ref{S t 1}) starting from
the SOS expression (Eq.~\ref{S 1 subst}).

\begin{figure}[ptbp]
\begin{center}
\includegraphics[clip=true, viewport=0.0in 3.0in 6.0in
10.0in,scale=0.6,angle=-90]{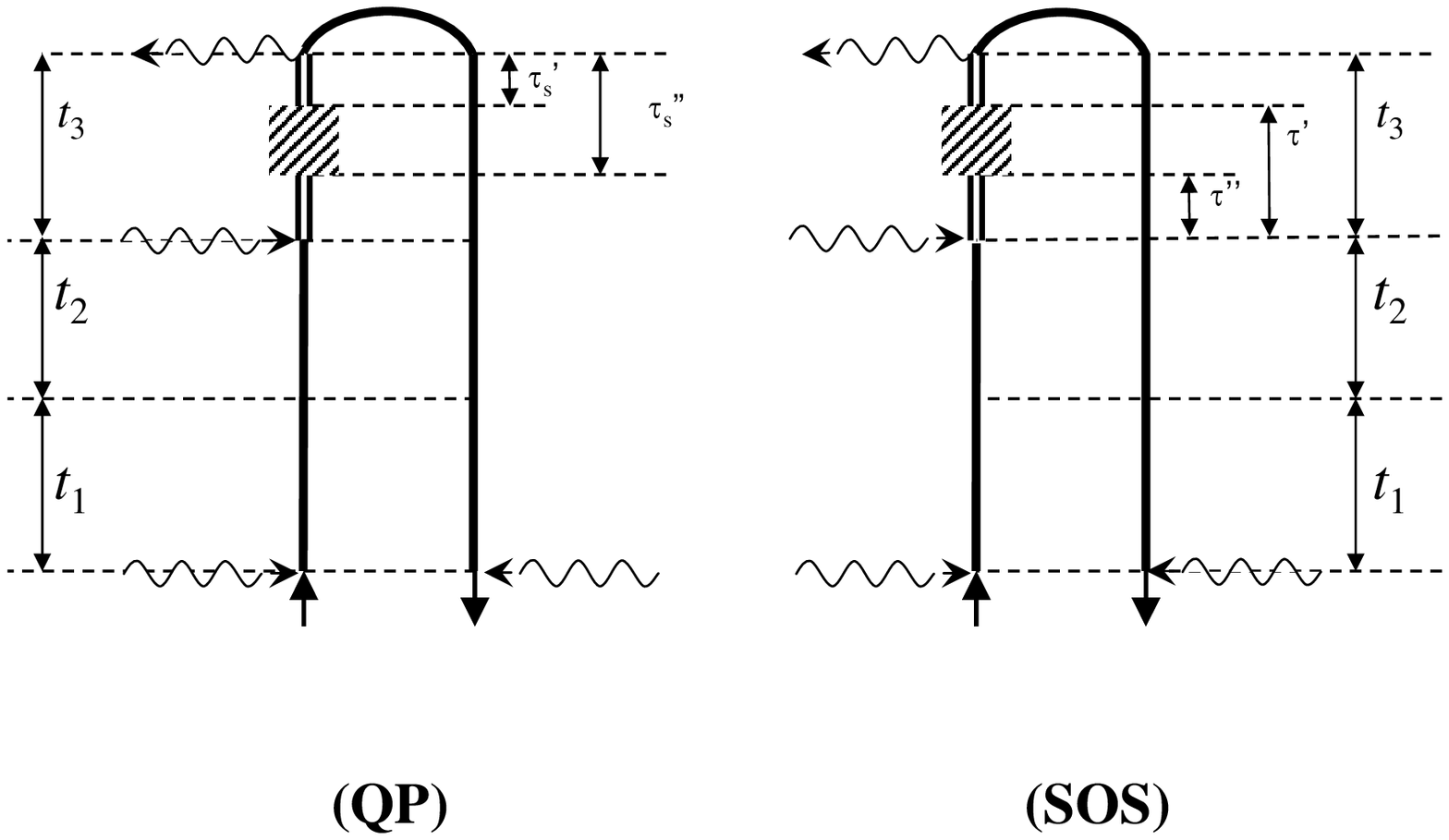}
\end{center}
\caption{Loop diagrams showing the order of time variables in the QP (Eq.~%
\protect\ref{S t 2}) and SOS (Eq.~\protect\ref{S 2 subst 2}) expressions for 
$S_{II}$.}
\label{Darius_K2}
\end{figure}

$S_{II}$ can be calculated similarly. By combining Eqs.~(\ref{S 2 subst})
and (\ref{B S time}) the same type of cancellation of harmonic terms yields%
\begin{align}
\bm{S}_{II}^{\left( SOS\right) } & =-\theta\left( t_{3}\right) \theta\left(
t_{2}\right) \theta\left( t_{1}\right)
\int_{0}^{t_{3}}d\tau^{\prime}\int_{0}^{\tau^{\prime}}d\tau^{\prime\prime}%
\times  \label{S 2 subst 2} \\
& \left\langle \hat{\bm{\mu}}^{-}\hat{G}^{\dagger}\left( t_{2}+t_{3}\right) 
\hat{\bm{\mu}}^{-}\mathcal{\hat{G}}_{0}\left( t_{3}-\tau^{\prime}\right)
\Gamma\left( \tau^{\prime}-\tau^{\prime\prime }\right) \mathcal{\hat{G}}%
_{0}\left( \tau^{\prime\prime}\right) \hat{\bm{\mu}}^{+}\hat{G}(t_{1}+t_{2})%
\hat{\bm{\mu}}^{+}\right\rangle .  \notag
\end{align}
Eq.~(\ref{S 2 subst 2}) is identical to Eq.~(\ref{S t 2}) as illustrated in
Fig.~(\ref{Darius_K2}).

We finally turn to $\bm{S}_{III}$, (Eq.~\ref{S 3 subst}). Using again the
Bethe Salpeter equation (\ref{B S time}) and the fact that terms that only
depend on $\mathcal{\hat{G}}_{0}$ must cancel (harmonic reference), we get%
\begin{align}
& \bm{S}_{III}^{\left( SOS\right) }=-\theta\left( t_{3}\right) \theta\left(
t_{2}\right) \theta\left( t_{1}\right) \times  \label{S 3 subst 2} \\
& \left[ \int_{0}^{t_{2}}d\tau^{\prime}\int_{0}^{\tau^{\prime}}d\tau
^{\prime\prime}\left\langle \hat{\bm{\mu}}^{-}\hat{G}(t_{3})\hat{\bm{\mu}}%
^{-}\mathcal{\hat{G}}_{0}(t_{2}-\tau^{\prime})\Gamma(\tau^{\prime}-\tau^{%
\prime\prime})\mathcal{\hat{G}}_{0}(\tau ^{\prime\prime})\hat{\bm{\mu}}^{+}%
\hat{G}(t_{1})\hat{\bm{\mu}}^{+}\right\rangle \right.  \notag \\
& \left.
+\int_{0}^{t_{2}+t_{3}}d\tau^{\prime}\int_{0}^{\tau^{\prime}}d\tau^{\prime%
\prime}\left\langle \hat{\bm{\mu}}^{-}\hat{G}^{\dagger }(t_{3})\hat{\bm{\mu}}%
^{-}\mathcal{\hat{G}}_{0}(t_{2}+t_{3}-\tau^{\prime})\Gamma(\tau^{\prime}-%
\tau^{\prime\prime})\mathcal{\hat{G}}_{0}(\tau^{\prime\prime})\hat{\bm{\mu}}%
^{+}\hat{G}(t_{1})\hat{\bm{\mu}}^{+}\right\rangle \right] .  \notag
\end{align}
The equivalence of QP (Eq.~\ref{S t 3}) and SOS (Eq.~\ref{S 3 subst})
expressions can be shown as follows: the two terms in Eq.~(\ref{S 3 subst 2}%
) are labeled (SOSa) and (SOSb). The term (SOSb) can further be split into
two terms (SOSb1) and (SOSb2), the first corresponding to $%
\tau^{\prime}<t_{2}$, the second to $\tau^{\prime}>t_{2}$ (Fig.~\ref%
{Darius_K3}). (SOSb1) is identical to (SOSa), but with opposite sign coming
from $\hat{G}^{\dagger }(t_{3})$. Only the second term (SOSb2) remains, and
it is equivalent to the (QP) diagram. We thus obtained Eq.~(\ref{S t 3})
from Eq.~(\ref{S 3 subst 2}).

\begin{figure}[ptb]
\includegraphics[clip=true, viewport=0.0in 0.0in 6.0in
11.0in,scale=0.6,angle=-90]{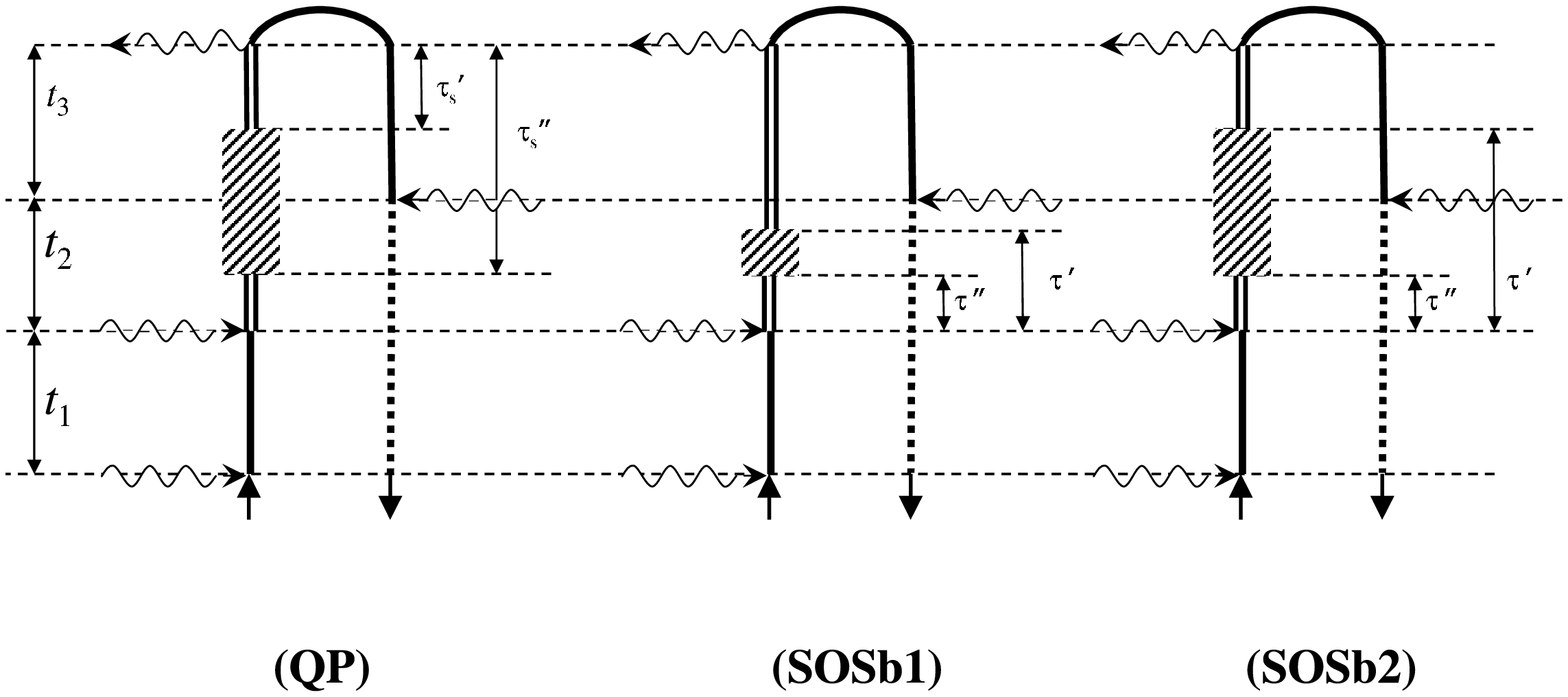} 
\caption{Loop diagrams showing the order of time variables in the QP (Eq.~%
\protect\ref{S t 3}) and SOS (Eq.~\protect\ref{S 3 subst 2}) expressions for 
$S_{III}$. (SOSb1) cancels with the (SOSa) term in (Eq.~\protect\ref{S 3
subst 2}) (not shown). The remaining diagram (SOSb2) is identical to the
(QP) diagram with a simple change of time variables.}
\label{Darius_K3}
\end{figure}

\section{2D correlation signals\label{Sec:2D}}

2D signals are displayed as correlation plots obtained by the double Fourier
transforms of the various signals.\cite{mukamel_annrev2000} We shall denote
the frequencies conjugate to the pulse delay times $t_{1},t_{2}$ and $t_{3}$
by $\Omega_{1},$ $\Omega_{2}$ and $\Omega_{3}$. Starting with Eq.~(\ref%
{eqn:Polarization3B}), and deleting some inessential factors, we obtain the
induced polarization, which depends parametrically on the delay times
$t_1$, $t_2$ and $t_3$:%
\begin{equation}
\bm{P}_{s}(t_3,t_2,t_1)=\bm{S}_{s}(t_{3},t_{2},t_{1})e^{i(\lambda_{1}%
\omega_{1}+\lambda_{2}\omega_{2}+\lambda_{3}\omega_{3})t_{3}}e^{i(%
\lambda_{1}\omega_{1}+\lambda_{2}\omega_{2})t_{2}}e^{i\lambda_{1}%
\omega_{1}t_{1}}.   \label{eqn:pol_carrier}
\end{equation}
Specifying the three possible signals by a proper choice of $\lambda$
factors we obtain:%
\begin{align*}
\bm{P}_{I}(t_3,t_2,t_1) & =\bm{S}_{I}(t_{3},t_{2},t_{1})e^{i(-\omega_{1}+%
\omega_{2}+\omega_{3})t_{3}}e^{i(-\omega_{1}+\omega_{2})t_{2}}e^{-i%
\omega_{1}t_{1}}, \\
\bm{P}_{II}(t_3,t_2,t_1) & =\bm{S}_{II}(t_{3},t_{2},t_{1})e^{i(\omega_{1}-%
\omega_{2}+\omega_{3})t_{3}}e^{i(\omega_{1}-\omega
_{2})t_{2}}e^{i\omega_{1}t_{1}}, \\
\bm{P}_{III}(t_3,t_2,t_1) & =\bm{S}_{III}(t_{3},t_{2},t_{1})e^{i(\omega_{1}+%
\omega_{2}-\omega_{3})t_{3}}e^{i(\omega_{1}+\omega
_{2})t_{2}}e^{i\omega_{1}t_{1}}.
\end{align*}
The 2DCS for $\bm{P}_{I}$ and $\bm{P}_{II}$ is defined as%
\begin{equation}
\bm{P}_{\alpha}(\Omega_{3},t_{2},\Omega_{1})\equiv\int_{0}^{\infty
}dt_{3}\int_{0}^{\infty}dt_{1}\bm{P}_{\alpha}(t_{3},t_{2},t_{1})\exp\left\{
i\Omega_{3}t_{3}+i\Omega_{1}t_{1}\right\} ,~~\alpha=I,II 
\label{eqn:pol alpha}
\end{equation}
For the SOS picture we use the expansions in eigenstates given by Eqs.~(\ref%
{eigen S1}), (\ref{eigen S2}) and (\ref{eigen S3}). The QP expressions for $%
\bm{P}_{I}^{\left( QP\right) }$, $\bm{P}_{II}^{\left( QP\right) }$ and $%
\bm{P}_{III}^{\left( QP\right) }$ are obtained along the lines presented in
App.~\ref{Resp fun}.
Dephasing  is introduced phenomenologically by adding a decay rate $\gamma$
to the Green's functions.
We thus obtain $\bm{P}_{I}$%
\begin{align}
& \bm{P}_{I}^{(SOS)}(\Omega_{3},t_{2},\Omega_{1})=  \label{eqn:SOS I freq} \\
& i\sum_{e_{2},e_{1}}\bm{\mu}_{ge_{1}}\bm{\mu}_{ge_{1}}^{\ast}\bm{\mu}%
_{ge_{2}}^{\ast}\bm{\mu}_{ge_{2}}I_{e_{2}}^{\ast}(-\Omega_{1}+%
\omega_{1})I_{g}^{\ast}(t_{2})I_{g}(t_{2})I_{e_{1}}(\Omega_{3}-\omega_{1}+%
\omega_{2}+\omega_{3})  \notag \\
& +i\sum_{e_{2},e_{1}}\bm{\mu}_{ge_{1}}\bm{\mu}_{ge_{2}}^{\ast}\bm{\mu}%
_{ge_{1}}^{\ast}\bm{\mu}_{ge_{2}}I_{e_{2}}^{\ast}(-\Omega_{1}+%
\omega_{1})I_{e_{2}}^{\ast}(t_{2})I_{e_{1}}(t_{2})I_{e_{1}}(\Omega_{3}-%
\omega_{1}+\omega_{2}+\omega_{3})  \notag \\
& -i\sum_{e_{2},e_{1}f}\bm{\mu}_{e_{2}f}\bm{\mu}_{e_{1}f}^{\ast}\bm{\mu}%
_{ge_{1}}^{\ast}\bm{\mu}_{ge_{2}}I_{e_{2}}^{\ast}(-\Omega_{1}+%
\omega_{1})I_{e_{2}}^{\ast}(t_{2})I_{e_{1}}(t_{2})\mathcal{F}%
_{fe_{2}}(\Omega_{3}-\omega_{1}+\omega_{2}+\omega _{3}),  \notag
\end{align}%
\begin{align}
& \bm{P}_{I}^{\left( QP\right) }(\Omega_{3},t_{2},\Omega _{1})=
\label{S 1 final} \\
& -2\sum_{e_{1},e_{2},e_{3},e_{4}}\bm{\mu}_{e_{4}}\bm{\mu }_{e_{3}}^{\ast}%
\bm{\mu}_{e_{2}}^{\ast}\bm{\mu}_{e_{1}}I_{e_{1}}^{%
\ast}(t_{2})I_{e_{2}}(t_{2})I_{e_{1}}^{\ast}(-\Omega_{1}-%
\omega_{1})I_{e_{4}}(\Omega_{3}-\omega_{1}+\omega_{2}+\omega_{3})  \notag \\
&
\times\Gamma_{e_{4}e_{1}e_{3}e_{2}}(\Omega_{3}-\omega_{1}+\omega_{2}+%
\omega_{3}+\varepsilon_{e_{1}}+i\gamma_{e_1})\mathcal{G}_{0\,e_{3}e_{2}}(%
\Omega_{3}-\omega_{1}+\omega_{2}+\omega_{3}+\varepsilon_{e_{1}}+i\gamma
_{e_1}).  \notag
\end{align}
The Green's function Fourier transform is defined as $G(\omega)=\int
dt\exp(i\omega t)G(t)$ {[}and $G(t)=\int\frac{d\omega}{2\pi}\exp(-i\omega
t)G(\omega)${]}. We have%
\begin{align}
I_{e}(\omega) & \equiv\left\langle e\left\vert \hat{G}\left( \omega\right)
\right\vert e\right\rangle =(\omega-\varepsilon_{e}+i\gamma_{e})^{-1},
\label{eqn:I omega} \\
\mathcal{G}_{0\,e_{2}e_{1}}(\omega) & \equiv\left\langle
e_{1}e_{2}\left\vert \mathcal{\hat{G}}_{0}(\omega)\right\vert
e_{1}e_{2}\right\rangle =\frac{1}{\omega-\varepsilon_{e_{2}}-%
\varepsilon_{e_{1}}+i\left( \gamma_{e_{2}}+\gamma_{e_{1}}\right) }. 
\label{eqn:G omega}
\end{align}
Eq.~(\ref{eqn:G omega}) is obtained by transforming $\mathcal{G}%
_{0\,kljr}\left( t\right) $ to the single-exciton basis and performing the
Fourier transform. We also define 
\begin{align*}
\mathcal{F}_{ab}(t) & \equiv-i\theta(t)\exp\left( i\left( \varepsilon
_{b}-\varepsilon_{a}\right) t-\left( \gamma_{a}+\gamma_{b}\right) t\right) ,
\\
\mathcal{F}_{ab}(\omega) &
=(\omega-\varepsilon_{a}+\varepsilon_{b}+i\gamma_{a}+i\gamma_{b})^{-1}.
\end{align*}
Similarly we obtain for $\bm{P}_{II}$:%
\begin{align}
& \bm{P}_{II}^{(SOS)}(\Omega_{3},t_{2},\Omega_{1})=  \label{eqn:SOS II freq}
\\
& -i\sum_{e_{2},e_{1}}\bm{\mu}_{ge_{2}}^{\ast}\bm{\mu }_{ge_{2}}\bm{\mu}%
_{ge_{1}}\bm{\mu}_{ge_{1}}^{\ast}I_{e_{1}}(\Omega_{1}+\omega_{1})I_{g}^{%
\ast}(t_{2})I_{g}(t_{2})I_{e_{2}}(\Omega
_{3}+\omega_{1}-\omega_{2}+\omega_{3})  \notag \\
& -i\sum_{e_{2},e_{1}}\bm{\mu}_{ge_{1}}\bm{\mu}_{ge_{2}}^{\ast}\bm{\mu}%
_{ge_{2}}\bm{\mu}_{ge_{1}}^{\ast}I_{e_{1}}(\Omega_{1}+\omega_{1})I_{e_{2}}^{%
\ast}(t_{2})I_{e_{1}}(t_{2})I_{e_{1}}(\Omega_{3}+\omega_{1}-\omega_{2}+%
\omega_{3})  \notag \\
& +i\sum_{e_{2},e_{1},f}\bm{\mu}_{e_{_{2}}f}\bm{\mu}_{e_{_{1}}f}^{\ast}%
\bm{\mu}_{ge_{_{2}}}\bm{\mu}_{ge_{_{1}}}^{\ast}I_{e_{1}}(\Omega_{1}+%
\omega_{1})I_{e_{2}}^{\ast}(t_{2})I_{e_{1}}(t_{2})\mathcal{F}%
_{fe_{2}}(\Omega_{3}+\omega_{1}-\omega_{2}+\omega _{3}),  \notag
\end{align}%
\begin{align}
& \bm{P}_{II}^{\left( QP\right) }(\Omega_{3},t_{2},\Omega _{1})=
\label{S 2 final} \\
& -2\sum_{e_{4}..e_{1}}\bm{\mu}_{e_{4}}\bm{\mu}_{e_{3}}^{\ast}\bm{\mu}%
_{e_{2}}\bm{\mu}_{e_{1}}^{\ast}I_{e_{2}}^{\ast
}(t_{2})I_{e_{1}}(t_{2})I_{e_{1}}(\Omega_{1}+\omega_{1})I_{e_{4}}(\Omega
_{3}+\omega_{1}-\omega_{2}+\omega_{3})  \notag \\
&
\times\Gamma_{e_{4}e_{2}e_{3}e_{1}}(\Omega_{3}+\omega_{1}-\omega_{2}+%
\omega_{3}+\varepsilon_{e_{2}}+i\gamma_{e_{2}})\mathcal{G}%
_{0\,e_{3}e_{1}}(\Omega_{3}+\omega_{1}-\omega_{2}+\omega_{3}+%
\varepsilon_{e_{2}}+i\gamma_{e_{2}}).  \notag
\end{align}
The $\bm{P}_{III}$ 2DCS signal is defined as 
\begin{equation}
\bm{P}_{III}(\Omega_{3},\Omega_{2},t_{1})\equiv\int_{0}^{\infty}dt_{3}%
\int_{0}^{\infty}dt_{2}\bm{P}_{III}(t_{3},t_{2},t_{1})\exp\left\{
i\Omega_{3}t_{3}+i\Omega_{2}t_{2}\right\} .   \label{eqn:pol alpha 2}
\end{equation}
This yields:%
\begin{align}
& \bm{P}_{III}^{(SOS)}(\Omega_{3},\Omega_{2},t_{1})=
\label{eqn:SOS III freq} \\
& -i\sum_{e_{2},e_{1},f}\bm{\mu}_{ge_{1}}\bm{\mu}_{e_{1}f}\bm{\mu}%
_{e_{2}f}^{\ast}\bm{\mu}_{ge_{2}}^{\ast}I_{e_{2}}(t_{1})I_{f}(\Omega_{2}+%
\omega_{1}+\omega_{2})I_{e_{1}}(\Omega_{3}+\omega _{1}+\omega_{2}-\omega_{3})
\notag \\
& +i\sum_{e_{2},e_{1},f}\bm{\mu}_{ge_{1}}\bm{\mu}_{e_{1}f}\bm{\mu}%
_{e_{2}f}^{\ast}\bm{\mu}_{ge_{2}}^{\ast}I_{e_{2}}(t_{1})I_{f}(\Omega_{2}+%
\omega_{1}+\omega_{2})\mathcal{F}_{fe_{1}}(\Omega
_{3}+\omega_{1}+\omega_{2}-\omega_{3}),  \notag
\end{align}%
\begin{align}
& \bm{P}_{^{III}}^{\left( QP\right) }(\Omega_{3},\Omega_{2},t_{1})=
\label{S 3 final} \\
& -2\sum_{e_{4}...e_{1}}\bm{\mu}_{e_{4}}\bm{\mu}_{e_{3}}\bm{\mu}%
_{e_{2}}^{\ast}\bm{\mu}_{e_{1}}^{\ast}I_{e_{1}}(t_{1})I_{e_{4}}(\Omega_{3}+%
\omega_{1}+\omega_{2}-\omega_{3})I_{e_{3}}^{\ast
}(\Omega_{2}-\Omega_{3}+\omega_{3})\times  \notag \\
& \left[ \Gamma_{e_{4}e_{3}e_{2}e_{1}}(\Omega_{2}+\omega_{1}+\omega _{2})%
\mathcal{G}_{0\,e_{2}e_{1}}(\Omega_{2}+\omega_{1}+\omega_{2})\right.  \notag
\\
& \left. -\Gamma_{e_{4}e_{3}e_{2}e_{1}}(\Omega_{3}+\omega_{1}+\omega
_{2}-\omega_{3}+\varepsilon_{e_{3}}+i\gamma_{e_{3}})\mathcal{G}%
_{0\,e_{2}e_{1}}(\Omega_{3}+\omega_{1}+\omega_{2}-\omega_{3}+%
\varepsilon_{e_{3}}+i\gamma_{e_{3}})\right] .  \notag
\end{align}

Both $\bm{P}^{(SOS)}$ and $\bm{P}^{\left( QP\right) }$ depend on the
single-exciton energies. However, the SOS expressions contain two-exciton
eigenenergies ($\varepsilon_f$) explicitly, while the QP counterparts contain the
scattering
matrix $\Gamma$ instead. The equivalence of the two representations has been
established in Sec.~\ref{Sec:Connection}. 
Eqs.~(\ref{eqn:SOS I freq}-\ref{S 3 final}) constitute our final expressions for
the various 2DCS signals. In this form they may be readily used in 
numerical simulations. The SOS expressions (Eqs.~\ref%
{eqn:SOS I freq}, \ref{eqn:SOS II freq} and \ref{eqn:SOS III freq}) were
recently used to survey the various possible resonances and cross-peaks
in 2DCS of semiconductors.%
\cite{Lijun2006}

\section{Discussion\label{Sec:Discuss}}

The quasiparticle representation is obtained using the Heisenberg equations
for the exciton oscillator variables. These equations form an infinite
hierarchy involving successively higher numbers of excitons.\cite%
{AbramaviciusMukamel2006b,ChernyakZhangMukamel98} The hierarchy may be
truncated, depending on the observable of interest. For instance, the
absorption originates from single-exciton creation/annihilation. Only
single-exciton variables should then be considered, and exciton-exciton
interaction terms may be neglected. The nonlinear response depends on the
exciton interactions, thus single- and double-exciton variables need to be
treated explicitly. The two coupled NEE equations (\ref{EOM 1},
\ref{EOM 2}) describe the third order
response. These equations are exact in the absence of dephasing. 
When dephasing is included by adding linear coupling to a phonon bath, two
additional variables $\left\langle \hat{B}^{\dagger}\hat{B}\right\rangle$
and $\left\langle \hat{B}^{\dagger}\hat{B}\hat{B}\right\rangle$ must be
included in the NEE to describe the third order response.
\cite{ChernyakZhangMukamel98} Without dephasing these may be
factorized as  $\left\langle \hat{B}^{\dagger}\hat{B}\right\rangle=%
\left\langle \hat{B}^{\dagger}\right\rangle \left\langle\hat{B}\right\rangle$
and $\left\langle \hat{B}^{\dagger}\hat{B}\hat{B}\right\rangle=%
\left\langle \hat{B}^{\dagger}\right\rangle%
\left\langle \hat{B}\hat{B}\right\rangle$. We then recover the coherent limit
considered in this article. For some techniques the present equations
provide a good approximation even in the presence of dephasing.
$\left\langle \hat{B}^{\dagger}\hat{B}\right\rangle$ describes incoherent
exciton transport and is only relevant during $t_2$, while
$\left\langle \hat{B}^{\dagger}\hat{B}\hat{B}\right\rangle$ is generated
during $t_3$. It describes the optical coherence between one-exciton and two-exciton
manifolds, which are represented by $\left\langle \hat{B}\right\rangle$ and
$\left\langle \hat{B}\hat{B}\right\rangle$.

The quasiparticle approach
avoids the explicit calculation of multiple exciton states: their influence
is represented by the scattering matrix, which can be calculated provided
the exciton interactions are known.
We have shown how the quasiparticle expressions for the various third order
techniques, ordinarily derived by solving equations of motion, can be
obtained directly from the sum-over-states expressions by employing the
Bethe-Salpether equation. These expressions explicitly contain the
two-exciton Green's functions and have many interfering terms with large
cancellations,\cite{spanomuk89} which complicate their numerical
implementation. In the QP picture, on the other hand, these interference
effects are
built-in, considerably simplifying the expressions for the nonlinear
response.\cite{spanomuk91a}

The interpretation of 2DCS signals using the SOS expressions is
straightforward.\cite{Lijun2006} In the $\mathbf{k}_{I}$ technique 
one-exciton coherences are observed during $t_1$, and the coherences
between excitons and biexcitons are observed during $t_3$. Thus the 2DCS shows peaks along 
$\Omega_1$ and $\Omega_3$ corresponding to these resonances.
$\mathbf{k}_{III}$ shows
biexciton resonances along the $\Omega_{2}$ axis, this technique is known in
NMR as
double-quantum coherence. We have established the connection between the
SOS and the QP pictures by using time-ordering on the
Keldysh-Schwinger loop, which only maintains partial time ordering in real
(physical) time.

The Hamiltonian $\hat{H}$ (Eq.~\ref{main H}) can describe several
microscopic models other than the Wannier excitons considered here ($\hat{H}_{T}$%
). Vibrational excitations (soft-core bosons) and Frenkel excitons
(hard-core bosons, Paulions) in molecules can be mapped into the same model.
\cite{ZhangMukamel1999,AbramaviciusMukamel2006b,KuhnChernyak96} The equations
of
motion for these other systems are similar, but not identical, because of
the different commutation relations (QP statistics). Eq.~(\ref{comm rel})
provides a unified description for all of these systems, by specifying the
proper commutation rules:\cite{ChernyakMukamel1996,ChernyakZhangMukamel98}
for bosons $\mathcal{P}_{mnpq}=0$ and for Paulions $\mathcal{P}%
_{mnpq}=\delta_{mn}\delta_{mq}\delta_{np}$. These expressions for $\mathcal{P%
}$ may be substituted into our final expressions for the response functions,
where they only affect the exciton scattering matrix, which in the frequency
domain reads (App.~\ref{Gamma calc}) 
\begin{equation*}
\Gamma(\omega)=\left( \mathbb{I}-V\mathcal{G}_{0}(\omega)\right) ^{-1}V%
\mathcal{G}_{0}(\omega)\left( \mathbb{I}-\mathcal{P}\right) \mathcal{G}%
_{0}^{-1}(\omega)-\mathcal{P}\mathcal{G}_{0}^{-1}(\omega)\text{,}
\end{equation*}
where $V$ is given in Eq.~(\ref{eqn:anh def}), $\mathcal{G}_{0}$ is the free
two--exciton Green's function (App.~\ref{EOM}) and $\mathbb{I}$ is the
tetradic identity matrix. The nonlinearity of the system depends on QP
interactions as well as non-boson statistics; both enter through $\Gamma$.
For noninteracting bosons, where $U\equiv\mathcal{P}\equiv0$, $\Gamma$
vanishes and so does the nonlinear response. In Appendix~\ref{App_coup_osc}
we present $\Gamma$ for bosons and Paulions.

We have used the symmetry $P_{mnpq}=P_{mnqp}$ in our derivation.
Since the boson commutation relations are simpler than for Fermi or Pauli
operators, a considerable effort has been devoted to mapping the original
problem with complicated commutation relations into a boson picture.\cite%
{IlinskaiaIlinski1996,agranovichtosich} The resulting boson Hamiltonian
contains additional interactions which compensate for the statistics. For
instance, the Frenkel exciton Hamiltonian for Paulions may be mapped into an
anharmonic Hamiltonian of bosons with quartic couplings. Bosonization\cite%
{ChernyakMukamel1996} is very convenient for describing exciton scattering:
the response functions derived for bosons can be applied for arbitrary
operators, provided we modify the Hamiltonian and express it in terms of
boson operators.

\begin{acknowledgments}
We wish to thank Dr Igor V. Schweigert for valuable discussions. This
research was supported by the National Science Foundation Grant no.
CHE-0446555 and the National Institutes of Health 2RO1-GM59230-05.
\end{acknowledgments}

\newpage

\appendix

\section{Exciton representation of the two-band Hamiltonian for fermions 
\label{Hamiltonian}}

By construction, the Hamiltonians $H$ (Eq.~(\ref{main H})) and $\hat{H}_{T}$
(Eq.~(\ref{start H})) are equivalent only in the physically relevant space
of single and double excitations. This is sufficient to calculate the
response to third order in the field $E\left( t\right) $. $\hat{H}$ may be
constructed using the following rules:

\begin{itemize}
\item since the Hamiltonian (\ref{start H}) conserves the number of
excitons, it should only contain products with equal number of $\hat{B}%
^{\dagger}$ and $\hat{B}$ operators (except for the $H_{I}$ term, which does
change the number of excitons)

\item a term $\hat{B}_{a_{1}}^{\dagger}\hat{B}_{a_{2}}^{\dagger}\ldots\hat
{%
B}_{a_{p}}^{\dagger}\hat{B}_{b_{1}}\hat{B}_{b_{2}}\ldots\hat{B}_{b_{p}}$
gives zero when acting on states with less than $p$ excitations and only
affects manifolds with $p$ excitations and higher.
\end{itemize}

The parameters of $\hat{H}$ can be obtained as follows. First we note that
no constant term $k$ should be added to (\ref{main H}), since it would
yield: $\left\langle 0\left\vert k\right\vert 0\right\rangle \neq0$, while $%
\left\langle 0\left\vert H_{T}\right\vert 0\right\rangle =0$. The $%
\sum_{m1,n2,k2,l1}W_{m_{1}n_{2}l_{1}k_{2}}c_{m_{1}}^{\dagger}d_{n_{2}}^{%
\dagger}d_{k_{2}}c_{l_{1}}$ term of $H_{T}$ can be written directly as $%
\sum_{m_{1},m_{2},n_{1},n_{2}}W_{m_{1}m_{2}n_{1}n_{2}}c_{m_{1}}^{\dagger
}d_{m_{2}}^{\dagger}d_{n2}c_{n1}=\sum_{m,n}W_{mn}\hat{B}_{m}^{\dagger}\hat
{%
B}_{n}$. Also the term describing the interaction with light can be obtained
directly. Using the second rule given above we immediately see that no terms
higher than $\hat{B}_{a_{1}}^{\dagger}\hat{B}_{a_{2}}^{\dagger}\hat{B}%
_{b_{1}}\hat{B}_{b_{2}}$ are necessary in the sub-space defined by functions 
$\left\vert 0\right\rangle $, $\hat{B}_{i}^{\dagger}\left\vert
0\right\rangle $ and $\hat{B}_{i}^{\dagger}\hat{B}_{j}^{\dagger}\left\vert
0\right\rangle $. We thus obtain the form given in (\ref{main H}). We next
calculate, in this sub-space, matrix elements of $\hat{H}$, and compare to
matrix elements of $\hat{H}_{T}$. In this way a one-to-one correspondence of
the parameters of $\hat{H}$ and $\hat{H}_{T}$ can be established.

Additional terms must be included in $\hat{H}$ in order to describe higher
order response functions. This can be done using the same rules.

\section{The Nonlinear Exciton Equations\label{EOM}}

The Heisenberg equation of motion (NEE) for the Hamiltonian (\ref{main H})
reads:%
\begin{align}
i\frac{d\left\langle \hat{B}_{n}\right\rangle }{dt} &
=\sum_{m}h_{nm}\left\langle \hat{B}_{m}\right\rangle -\bm{\mu}_{n}^{\ast }%
\bm{E}^{+}+\sum_{mpq}V_{nmpq}\left\langle \hat{B}_{m}^{\dagger}\hat{B}_{p}%
\hat{B}_{q}\right\rangle  \label{EOM main} \\
& +\bm{E}^{+}\sum_{mpq}\mathcal{P}_{nmpq}\left( \left\langle \hat {B}%
_{m}^{\dagger}\hat{B}_{p}\right\rangle \bm{\mu}_{q}^{\ast }+\left\langle 
\hat{B}_{m}^{\dagger}\hat{B}_{q}\right\rangle \bm{\mu }_{p}^{\ast}\right) 
\notag
\end{align}
Here we invoked RWA and used the notation of Eq.~(\ref{eqn:Field}).
Employing (\ref{P def}) we see that $\mathcal{P}_{nmpq}=\mathcal{P}_{nmqp}$,
so the last two terms in Eq.~(\ref{EOM main}) can be recast as: $2\bm{E}^{%
\bm{+}}\sum_{mpq}\mathcal{P}_{nmpq}\left\langle \hat{B}_{m}^{\dagger}\hat{B}%
_{q}\right\rangle \bm{\mu}_{p}^{\ast}$. We now make the following
factorization:%
\begin{equation}
\left\langle \hat{B}_{m}^{\dagger}\hat{B}_{p}\right\rangle=%
\left\langle \hat{B}_{m}^{\dagger}\right\rangle %
\left\langle \hat{B}_{p}\right\rangle \textrm{ and }
\left\langle \hat{B}_{m}^{\dagger}\hat{B}_{p}\hat{B}_{q}\right\rangle
=\left\langle \hat{B}_{m}^{\dagger}\right\rangle \left\langle \hat{B}_{p}%
\hat{B}_{q}\right\rangle ,   \label{qqq}
\end{equation}
which is exact for pure states when dephasing is
neglected\cite{leegwaterMukamel92}
and is a good approximation in the absence of incoherent exciton
transport.
Eq.~(\ref{EOM main}) then yields the Eqs.~(\ref{EOM 1})
and (\ref{EOM 2}), where%
\begin{equation}
h_{mn,kl}^{\left( Y\right)
}=\delta_{mk}h_{nl}+\delta_{nl}h_{mk}+V_{mnkl}\equiv\bar{h}+V. 
\label{h(Y) def}
\end{equation}
We next expand the EOMs in orders of $E$. Using $B_{m}^{\left( 1\right) }$
for $\left\langle \hat{B}_{m}\right\rangle ^{\left( 1\right) }$ we obtain:%
\begin{align*}
i\frac{dB_{m}^{\left( 1\right) }}{dt} & =\sum_{n}h_{mn}B_{n}^{\left(
1\right) }-\bm{\mu}_{m}^{\ast}\bm{E}^{\bm{+}}\left( t\right) , \\
i\frac{dY_{mn}^{\left( 2\right) }}{dt} & =\sum_{kl}h_{mn,kl}^{\left(
Y\right) }Y_{kl}^{\left( 2\right) }-\bm{E}^{\bm{+}}\left( t\right) \left(
B_{n}^{\left( 1\right) }\bm{\mu}_{m}^{\ast}+B_{m}^{\left( 1\right) }\bm{\mu}%
_{n}^{\ast}\right) +2\bm{E}^{\bm{+}}\left( t\right) \sum_{k,l}\mathcal{P}%
_{mnkl}B_{k}^{\left( 1\right) }\bm{\mu}_{l}^{\ast}, \\
i\frac{dB_{m}^{\left( 3\right) }}{dt} & =\sum_{n}h_{mn}B_{m}^{\left(
3\right) }+\sum_{nkl}V_{mnkl}B_{n}^{\left( 1\right) \ast}Y_{kl}^{\left(
2\right) }+2\bm{E}^{\bm{+}}\left( t\right) \sum _{npq}\mathcal{P}%
_{mnpq}B_{n}^{\left( 1\right) \ast}B_{q}^{\left( 1\right) }\bm{\mu}%
_{p}^{\ast}.
\end{align*}
The Green's function (tetradic matrix) for $Y^{\left( 2\right) }$ is $%
\mathcal{G}\left( t\right) _{mnkl}=-i\theta\left( t\right) \left[ \exp\left(
-ih^{\left( Y\right) }t\right) \right] _{mnkl}$, thus%
\begin{align*}
& Y_{mn}^{\left( 2\right) }\left( t\right) = \\
& -\int_{-\infty}^{\infty}\sum_{kl}\mathcal{G}_{mnkl}\left( t-\tau\right) %
\bm{E}^{\bm{+}}\left( \tau\right) \left[ \left( B_{l}^{\left( 1\right)
}\left( \tau\right) \bm{\mu}_{k}^{\ast }+B_{k}^{\left( 1\right) }\left(
\tau\right) \bm{\mu}_{l}^{\ast }\right) -2\sum_{pq}\mathcal{P}%
_{klpq}B_{p}^{\left( 1\right) }\left( \tau\right) \bm{\mu}_{q}^{\ast}\right]
d\tau \\
&
=+\int_{-\infty}^{\infty}d\tau\int_{-\infty}^{\infty}d\tau^{\prime}\sum_{kla}%
\mathcal{G}_{mnkl}\left( t-\tau\right) \times \\
& \left[ G_{la}\left( \tau-\tau^{\prime}\right) \bm{\mu}_{k}^{\ast}+G_{ka}%
\left( \tau-\tau^{\prime}\right) \bm{\mu}_{l}^{\ast }-2\sum_{pq}\mathcal{P}%
_{klpq}G_{pa}\left( \tau-\tau^{\prime}\right) \bm{\mu}_{q}^{\ast}\right] %
\bm{\mu}_{a}^{\ast}\bm{E}^{\bm{+}}\left( \tau\right) \bm{E}^{\bm{+}}\left(
\tau^{\prime}\right) .
\end{align*}
We also define the zero-order tetradic Green's function $\mathcal{G}_{0}$
for $Y^{\left( 2\right) }$ for the case $V=0$, i.e. $\mathcal{G}%
_{0\,mnkl}\left( t\right) =-i\theta\left( t\right) \left[ \exp\left( -i%
\bar
{h}t\right) \right] _{mnkl}$, it will be used later. $B_{m}^{\left(
3\right) }$ is given as%
\begin{align*}
B_{m}^{\left( 3\right) }\left( t^{\prime}\right) &
=\sum_{n}\int_{-\infty}^{\infty}G_{mn}\left( t^{\prime}-t\right) \times \\
& \left[ \sum_{pkl}V_{npkl}B_{p}^{\left( 1\right) \ast}\left( t\right)
Y_{kl}^{\left( 2\right) }\left( t\right) +2\bm{E}^{\bm{+}}\left( t\right)
\sum_{kpq}\mathcal{P}_{nkpq}B_{k}^{\left( 1\right) \ast }\left( t\right)
B_{q}^{\left( 1\right) }\left( t\right) \bm{\mu }_{p}^{\ast}\right] dt.
\end{align*}
This expression can be simplified using the symmetry $\mathcal{G}_{klfg}=%
\mathcal{G}_{klgf}$. At this point we introduce the tetradic exciton
scattering matrix $\Gamma$ defined as:%
\begin{equation}
\Gamma\left( \omega\right) \mathcal{G}_{0}\left( \omega\right) =V\mathcal{G}%
\left( \omega\right) \left( \mathbb{I}-\mathcal{P}\right) -\mathcal{P}, 
\label{Gamma def}
\end{equation}
which in time domain can be written as (see App. \ref{Gamma calc}):%
\begin{equation}
V\mathcal{G}\left( t-\tau\right) \left( \mathbb{I}-\mathcal{P}\right) =%
\mathcal{P}\delta\left( t-\tau\right)
+\int\limits_{-\infty}^{\infty}d\tau_{1}\Gamma\left( t-\tau_{1}\right) 
\mathcal{G}_{0}\left( \tau_{1}-\tau\right) ,   \label{Gamma def temp}
\end{equation}
where the tetradic identity matrix $\mathbb{I}_{fgjr}=\delta_{fj}\delta_{gr}.
$ Since $\mathcal{G}\left( t-\tau\right) \sim\theta\left( t-\tau\right) $ is
retarded, $\Gamma$ must be retarded as well, i.e., $\Gamma\left( t-\tau
_{1}\right) \sim\theta\left( t-\tau_{1}\right) $. To proceed further we take
advantage of the factorization:%
\begin{equation}
\mathcal{G}_{0\,kljr}\left( t\right) =iG_{kj}\left( t\right) G_{lr}\left(
t\right) ,   \label{F fact}
\end{equation}
which can be easily shown in the single-exciton eigenbasis. After a
rearrangement of terms we obtain:

\begin{align*}
B_{n4}^{\left( 3\right) }\left( \tau_{4}\right) & =-2\int\int\int \int\int
d\tau^{\prime}d\tau_{2}d\tau_{1}d\tau^{\prime\prime}d\tau_{3} \\
& \sum_{\substack{ n_{1}^{\prime},n_{2}^{\prime}, \\ n_{3}^{\prime},n_{4}^{%
\prime}}}\sum_{n_{1},n_{2},n_{3}}G_{n_{4}n_{4}^{\prime}}\left(
\tau_{4}-\tau^{\prime\prime}\right)
\Gamma_{n_{4}^{\prime}n_{3}^{\prime}n_{1}^{\prime}n_{2}^{\prime}}\left(
\tau^{\prime\prime}-\tau^{\prime}\right)
G_{n_{3}^{\prime}n_{3}}^{\ast}\left( \tau^{\prime\prime}-\tau_{3}\right)
\times \\
& G_{n_{2}^{\prime}n_{2}}\left( \tau^{\prime}-\tau_{2}\right)
G_{n_{1}^{\prime}n_{1}}\left( \tau^{\prime}-\tau_{1}\right) \theta\left(
\tau_{2}-\tau_{1}\right) \bm{\mu}_{n_{2}}^{\ast}\bm{\mu }_{n_{1}}^{\ast}%
\bm{\mu}_{n3}\bm{E}^{\bm{+}}\left( \tau_{2}\right) \bm{E}^{\bm{+}}\left(
\tau_{1}\right) \bm{E}^{\bm{-}}\left( \tau_{3}\right) .
\end{align*}
The 3rd order polarization is%
\begin{align*}
P^{\left( 3\right) }\left( \tau_{4}\right) & =\sum_{n_{4}}\left( \bm{\mu}%
_{n_{4}}B_{n_{4}}^{\left( 3\right) }+\bm{\mu}_{n_{4}}^{\ast}B_{n_{4}}^{%
\left( 3\right) \dagger}\right) \\
& =-2\int\int\int d\tau_{3}d\tau_{2}d\tau_{1}\sum_{n_{4},n_{3},n_{2},n_{1}}%
\bm{\mu}_{n_{4}}\bm{\mu}_{n3}\bm{\mu}_{n_{2}}^{\ast }\bm{\mu}_{n_{1}}^{\ast}
\\
& \theta\left( \tau_{2}-\tau_{1}\right) \int d\tau^{\prime\prime}\int
d\tau^{\prime}\sum_{n_{1}^{\prime},n_{2}^{\prime},n_{3}^{\prime},n_{4}^{%
\prime}}\Gamma_{n_{4}^{\prime}n_{3}^{\prime}n_{2}^{\prime}n_{1}^{\prime}}%
\left( \tau^{\prime\prime}-\tau^{\prime}\right) \times \\
& G_{n_{4}n_{4}^{\prime}}\left( \tau_{4}-\tau^{\prime\prime}\right)
G_{n_{3}^{\prime}n_{3}}^{\ast}\left( \tau^{\prime\prime}-\tau_{3}\right)
G_{n_{2}^{\prime}n_{2}}\left( \tau^{\prime}-\tau_{2}\right)
G_{n_{1}^{\prime}n_{1}}\left( \tau^{\prime}-\tau_{1}\right) \times \\
& \bm{E}^{\bm{-}}\left( \tau_{3}\right) \bm{E}^{\bm{+}}\left(
\tau_{2}\right) \bm{E}^{\bm{+}}\left( \tau_{1}\right) +\text{complex
conjugate},
\end{align*}
where we used $\Gamma_{n_{4}^{\prime}n_{3}^{\prime}n_{1}^{\prime}n_{2}^{%
\prime}}\left( t\right)
=\Gamma_{n_{4}^{\prime}n_{3}^{\prime}n_{2}^{\prime}n_{1}^{\prime}}\left(
t\right) $.

The above expression is finite only for $\tau_{2}>\tau_{1}.$ Hence there are
3 possible intervals for $\tau_{3},$ that define three contributions to the
third order response function $\bm{S}^{\left( QP\right) }$ 
\begin{align}
\bm{P}\left( \tau_{4}\right) & =\int\limits_{-\infty}^{\infty
}d\tau_{2}\int\limits_{-\infty}^{\infty}d\tau_{1}\theta\left(
\tau_{2}-\tau_{1}\right) \times  \label{eqn:contrib to P} \\
& \left( \int\limits_{-\infty}^{\tau_{1}}\bm{S}_{I}^{\left( QP\right)
}d\tau_{3}+\int\limits_{\tau_{1}}^{\tau_{2}}\bm{S}_{II}^{\left( QP\right)
}d\tau_{3}+\int\limits_{\tau_{2}}^{+\infty }\bm{S}_{III}^{\left( QP\right)
}d\tau_{3}\right) \bm{E}^{\bm{-}}\left( \tau_{3}\right) \bm{E}^{\bm{+}%
}\left( \tau_{2}\right) \bm{E}^{\bm{+}}\left( \tau _{1}\right) +c.c.  \notag
\end{align}
This definition of $\bm{S}_{I}^{\left( QP\right) }$, $\bm{S}_{II}^{\left(
QP\right) }$and $\bm{S}_{III}^{\left( QP\right) }$ is used in Appendix~(\ref%
{Resp fun}) to obtain the Eqs.~(\ref{S t 1}-\ref{S t 3}).

\section{Response functions of quasiparticles\label{Resp fun}}

For calculating each of the contributions to Eq.~(\ref{eqn:contrib to P}) we
need to switch to a different set of time-ordered variables. For $\bm{S}%
_{I}^{\left( QP\right) }$ we set: $\tau_{1}\rightarrow
\tau_{2},\tau_{2}\rightarrow\tau_{3},\tau_{3}\rightarrow\tau_{1}$:%
\begin{align*}
& \bm{S}_{I}^{\left( QP\right) }\left( \tau_{4},\tau_{3},\tau
_{2},\tau_{1}\right) =-2\sum_{n_{4},n_{3},n_{2},n_{1}}\bm{\mu}_{n_{4}}%
\bm{\mu}_{n3}\bm{\mu}_{n_{2}}^{\ast}\bm{\mu }_{n_{1}}^{\ast}\int_{-%
\infty}^{+\infty}d\tau^{\prime\prime}\int_{-\infty
}^{+\infty}d\tau^{\prime}\sum_{n_{1}^{\prime},n_{2}^{\prime},n_{3}^{\prime
},n_{4}^{\prime}}\times \\
& \Gamma_{n_{4}^{\prime}n_{3}^{\prime}n_{2}^{\prime}n_{1}^{\prime}}\left(
\tau^{\prime\prime}-\tau^{\prime}\right) G_{n_{4}n_{4}^{\prime}}\left(
\tau_{4}-\tau^{\prime\prime}\right) G_{n_{3}^{\prime}n_{3}}^{\ast}\left(
\tau^{\prime\prime}-\tau_{1}\right) G_{n_{2}^{\prime}n_{2}}\left(
\tau^{\prime}-\tau_{3}\right) G_{n_{1}^{\prime}n_{1}}\left( \tau^{\prime
}-\tau_{2}\right) ,
\end{align*}
\ where $\tau_{4}>\tau_{3}>\tau_{2}>\tau_{1}$.

Substituting $\tau_{s}^{\prime}=\tau_{4}-\tau^{\prime\prime}$ and $\tau
_{s}^{\prime\prime}=\tau_{4}-\tau^{\prime}$ and exchanging the dummy indices 
$n_{1}\rightarrow n_{2},n_{2}\rightarrow n_{3},n_{3}\rightarrow n_{1}$ (same
for primed indices) we obtain Eq.~(\ref{S t 1}). Integration limits for $%
\tau_{s}^{\prime}$ have been limited by $G_{n_{4}n_{4}^{\prime}}\left(
\tau_{s}^{\prime}\right) $ and $\Gamma\left(
\tau_{s}^{\prime\prime}-\tau_{s}^{\prime}\right) $, while for $%
\tau_{s}^{\prime\prime}$ by $G_{n_{3}^{\prime}n_{3}}\left(
\tau_{43}-\tau_{s}^{\prime\prime}\right) $. Eqs.~(\ref{S t 2}) and (\ref{S t
3}) are obtained in a similar way.

Eq.~(\ref{S t 1}) can be simplified considerably by performing the double
time-integrations analytically. We first express the exciton Green's
function $G\left( \tau\right) $ and $\Gamma\left( \tau\right) $ in the
one-exciton basis $\psi_{e}$, defined by:%
\begin{equation}
\sum_{n}h_{mn}\psi_{en}=\varepsilon_{e}\psi_{em}\text{,}   \label{exc basis}
\end{equation}
where $h_{mn}$ is given by Eq.~(\ref{int defin}). The energies $\varepsilon
_{e}$ define the lowest optically-excited manifold of the system, i.e.,
single excitons. In this basis we express the time and frequency-domain
one-exciton Green's functions:%
\begin{equation}
G_{mn}\left( \tau\right) =\sum_{e}\psi_{em}I_{e}\left( \tau\right)
\psi_{en}^{\ast}\Rightarrow I_{e}\left( \tau\right) =-i\theta\left(
\tau\right) \exp\left( -i\varepsilon_{e}t\right) ,   \label{I def}
\end{equation}
where we introduce dephasing via $\varepsilon\rightarrow\varepsilon
-i\gamma_{e}$. The tetradic exciton scattering matrix is given by%
\begin{equation*}
\Gamma_{m_{4}m_{3}m_{2}m_{1}}\left( \tau\right) =\sum_{e1\ldots
e4}\psi_{e_{4}m_{4}}\psi_{e_{3}m_{3}}\Gamma_{e_{4}e_{3}e_{2}e_{1}}\left(
\tau\right) \psi_{e_{2}m_{2}}^{\ast}\psi_{e_{1}m_{1}}^{\ast}\text{.}
\end{equation*}
and the transformed dipole matrix elements $\bm{\mu}_{eg}$ are $\bm{\mu}%
_{e}=\sum_{m}\bm{\mu}_{m}\psi_{em}$. Using these quantities we express the $%
\bm{S}_{I}^{\left( QP\right) }$ in the single-exciton basis:

\begin{align}
\bm{S}_{I}^{\left( QP\right) }(\tau_{4},\tau_{3},\tau_{2},\tau_{1}) &
=-2\theta\left( \tau_{43}\right) \theta\left( \tau_{32}\right) \theta\left(
\tau_{21}\right) \sum_{e_{1}\ldots e_{4}}\bm{\mu }_{e_{4}}\bm{\mu}%
_{e_{2}}^{\ast}\bm{\mu}_{e_{1}}^{\ast }\bm{\mu}_{e_{3}}  \label{s1_e} \\
&
\quad\quad\times\int_{0}^{\tau_{43}}d\tau_{s}^{\prime\prime}\int_{0}^{%
\tau_{s}^{\prime\prime}}d\tau_{s}^{\prime}\Gamma_{e_{4}e_{3}e_{2}e_{1}}(%
\tau_{s}^{\prime\prime}-\tau_{s}^{\prime})  \notag \\
& \quad\quad\times I_{e_{4}}(\tau_{s}^{\prime})I_{e_{2}}(\tau_{43}-\tau
_{s}^{\prime\prime})I_{e_{1}}(\tau_{42}-\tau_{s}^{\prime\prime})I_{e_{3}}^{%
\ast}(\tau_{41}-\tau_{s}^{\prime}).  \notag
\end{align}
We next introduce $\Gamma\left( t\right) =\int\frac{d\omega}{2\pi }%
e^{-i\omega t}\Gamma\left( \omega\right) $. Since the response function
depends only on the pulse delays and not on the absolute times, we denote $%
\bm{S}_{I}(\tau_{4},\tau_{3},\tau_{2},\tau_{1})=\bm{S}_{I}(t_{3},t_{2},t_{1})
$, where $t_{3}=\tau_{43}$, $t_{2}=\tau_{32}$ and $t_{1}=\tau_{21}$. We
perform a Fourier transform with respect to the first and last arguments. We
thus obtain%
\begin{align}
\bm{S}_{I}^{\left( QP\right) }(\Omega_{3},t_{2},\Omega_{1}) &
=-2i\theta\left( t_{2}\right) \sum_{e1\ldots e4}\bm{\mu}_{e_{4}}\bm{\mu}%
_{e_{3}}^{\ast}\bm{\mu}_{e_{2}}^{\ast}\bm{\mu }_{e_{1}}I_{e_{1}}^{\ast}%
\left( -\Omega_{1}\right) I_{e_{1}}^{\ast}\left( t_{2}\right)
I_{e_{_{2}}}\left( t_{2}\right) I_{e_{_{4}}}\left( \Omega _{3}\right) \times
\label{be4 Cauchy} \\
& \frac{1}{2\pi}\int d\omega\Gamma_{e_{_{4}}e_{_{1}}e_{_{3}}e_{_{2}}}\left(
\omega\right) \mathcal{G}_{0\,e_{_{3}}e_{_{2}}}\left( \omega\right)
I_{e_{1}}^{\ast}\left( \omega-\Omega_{3}\right) ,  \notag
\end{align}
The $\omega$ integration can be performed by noting that 
\begin{equation*}
\Gamma(\omega)\mathcal{G}_{0}(\omega)\sim\frac{1}{\omega-2\varepsilon
+2i\gamma}~, 
\end{equation*}
which is obtained from Eq.~(\ref{Gamma def}) by noting that $\mathcal{G}%
\left( \omega\right) $ has poles only at two-exciton energies, and that $%
2\varepsilon-2i\gamma$ is a good approximation for two-exciton energy and
dephasing rate. Hence, if we close the Cauchy integration path in the
positive half-plane, there will be only a single pole at $%
\omega=\Omega_{3}+\varepsilon_{e_{1}}+i\gamma_{e_{1}}$ as seen from (\ref%
{eqn:I omega}). This finally gives%
\begin{align}
\bm{S}_{I}^{\left( QP\right) }(\Omega_{3},t_{2},\Omega_{1}) &
=-2\sum_{e_{4}...e_{1}}\bm{\mu}_{e_{4}}\bm{\mu}_{e_{3}}^{\ast }\bm{\mu}%
_{e_{2}}^{\ast}\bm{\mu}_{e_{1}}I_{e_{1}}^{%
\ast}(t_{2})I_{e_{2}}(t_{2})I_{e_{1}}^{\ast}(-\Omega_{1})I_{e_{4}}(\Omega
_{3})  \label{S 1 simple} \\
&
\times\Gamma_{e_{4}e_{1}e_{3}e_{2}}(\Omega_{3}+\varepsilon_{e_{1}}+i%
\gamma_{1})\mathcal{G}_{0\,e_{3}e_{2}}(\Omega_{3}+\varepsilon_{e_{1}}+i%
\gamma_{1}),  \notag
\end{align}
To account for carrier frequencies $\omega_{1}$, $\omega_{2}$ and $\omega_{3}
$ appearing in the polarization (Eq.~\ref{eqn:pol_carrier}) at this stage,
we can perform the substitution $\Omega_{1}\rightarrow\Omega_{1}+\omega_{1}$
and $\Omega_{3}\rightarrow-\Omega_{1}+\omega_{1}+\omega_{2}+\omega_{3}$. In
this way we obtain Eq.~(\ref{S 1 final}). Eqs.~(\ref{S 2 final}) and (\ref{S
3 final}) are derived similarly.

\section{The exciton scattering-matrix\label{Gamma calc}}

In order to use equations (\ref{S t 1}-\ref{S t 3}) for calculating the
quasiparticle response function, we should calculate the scattering matrix $%
\Gamma$. We first write $\mathcal{G}\left( \omega\right) $ and $\mathcal{G}%
_{0}\left( \omega\right) $ in an operator form%
\begin{align*}
\mathcal{G}_{0}\left( \omega\right) & =\frac{1}{\omega-\bar{h}} \\
\mathcal{G}\left( \omega\right) & =\frac{1}{\omega-h^{\left( Y\right) }}=%
\frac{1}{\omega-\bar{h}-V}
\end{align*}
where $\bar{h}$ is defined in (\ref{h(Y) def}). The Dyson equation then reads%
\begin{equation*}
\mathcal{G}=\mathcal{G}_{0}+\mathcal{G}_{0}V\mathcal{G}\left( \omega\right) =%
\mathcal{G}_{0}+\mathcal{G}_{0}V\mathcal{G}_{0}+\mathcal{G}_{0}V\mathcal{G}%
_{0}V\mathcal{G}_{0}+\ldots, 
\end{equation*}
which can be recast in the form%
\begin{equation*}
V\mathcal{G}=\left( 1-V\mathcal{G}_{0}\right) ^{-1}V\mathcal{G}_{0}. 
\end{equation*}
Using Eq.~(\ref{Gamma def}), we obtain:%
\begin{equation*}
\Gamma\mathcal{G}_{0}=\left( \mathbb{I}-V\mathcal{G}_{0}\right) ^{-1}V%
\mathcal{G}_{0}\left( \mathbb{I}-\mathcal{P}\right) -\mathcal{P}, 
\end{equation*}
which results in the final expression for $\Gamma$%
\begin{equation}
\Gamma=\left( \mathbb{I}-V\mathcal{G}_{0}\right) ^{-1}V\mathcal{G}_{0}\left( 
\mathbb{I}-\mathcal{P}\right) \mathcal{G}_{0}^{-1}-\mathcal{P}\mathcal{G}%
_{0}^{-1}   \label{eqn:Gamma_Final}
\end{equation}
The l.h.s. of Eq.~(\ref{Gamma def}) can be expressed as a convolution: 
\begin{equation*}
\int d\tau^{\prime}\int dt_{1}\Gamma\left( t_{1}\right) \mathcal{G}%
_{0}\left( \tau^{\prime}-t_{1}\right) \exp\left( i\omega\tau^{\prime
}\right) . 
\end{equation*}
The r.h.s. can be written as%
\begin{equation*}
V\int d\tau^{\prime}\mathcal{G}\left( \tau^{\prime}\right) \left( \mathbb{I}-%
\mathcal{P}\right) \exp\left( i\omega\tau^{\prime}\right) -\int
d\tau^{\prime}\mathcal{P\delta}\left( \tau^{\prime}\right) \exp\left(
i\omega\tau^{\prime}\right) , 
\end{equation*}
note that $\mathcal{P}$ is independent on $\tau^{\prime}$ or $\omega$. Since
l.h.s.=r.h.s. for any $\omega$, we must have:%
\begin{equation*}
\int dt_{1}\Gamma\left( t_{1}\right) \mathcal{G}_{0}\left( \tau^{\prime
}-t_{1}\right) =V\mathcal{G}\left( \tau^{\prime}\right) \left( \mathbb{I}-%
\mathcal{P}\right) -\mathcal{P\delta}\left( \tau^{\prime}\right) . 
\end{equation*}
Substituting $\tau^{\prime}\rightarrow t-\tau$ and $t_{1}\rightarrow
t-\tau_{1}$ we obtain Eq.~(\ref{Gamma def temp}).

\section{SOS expressions for third order techniques.\label{I_expressions}}

Upon expansion in the eigenstates for the exciton level scheme shown in Fig.~%
\ref{fig:exciton-model} we get 
\begin{align*}
\left\langle \hat{\bm{\mu}}^{-}\hat{\bm{\mu}}^{\bm{+}}\hat{\bm{\mu}}^{-}\hat{%
\bm{\mu}}^{+}\right\rangle & =\sum_{e,e^{\prime}}\left\langle \bm{\mu}%
_{ge^{\prime}}^{-}\bm{\mu}_{e^{\prime}g}^{+}\bm{\mu}_{ge}^{-}\bm{\mu }%
_{eg}^{+}\right\rangle , \\
\left\langle \hat{\bm{\mu}}^{-}\hat{\bm{\mu}}^{-}\hat{\bm{\mu}}^{+}\hat{%
\bm{\mu}}^{+}\right\rangle & =\sum_{e,e^{\prime}}\sum_{f}\left\langle %
\bm{\mu}_{ge^{\prime}}^{-}\bm{\mu}_{e^{\prime}f}^{-}\bm{\mu}_{fe}^{\bm{+}}%
\bm{\mu}_{eg}^{+}\right\rangle .
\end{align*}

Expanding Eqs.~(\ref{tech I} - \ref{tech III}) in the eigenstates, we obtain
the sum-over-states expressions for the third-order response functions:%
\begin{align}
\bm{S}_{I}^{(SOS)}(t_{3},t_{2},t_{1}) &
=i^{3}\theta(t_{3})\theta(t_{2})\theta(t_{1})\sum_{e,e^{\prime}}\bm{\mu}%
_{ge^{\prime}}\bm{\mu}_{e^{\prime}g}\bm{\mu}_{ge}\bm{\mu}_{eg}I_{e^{%
\prime}}^{\ast}\left( t_{1}\right) I_{e}\left( t_{3}\right)  \label{eigen S1}
\\
& +i^{3}\theta(t_{3})\theta(t_{2})\theta(t_{1})\sum_{e,e^{\prime}}\bm{\mu}%
_{ge^{\prime}}\bm{\mu}_{e^{\prime}g}\bm{\mu }_{ge}\bm{\mu}%
_{eg}I_{e^{\prime}}^{\ast}\left( t_{2}+t_{1}\right) I_{e}\left(
t_{2}+t_{3}\right)  \notag \\
& -i^{3}\theta(t_{3})\theta(t_{2})\theta(t_{1})\sum_{e,e^{\prime}}\sum _{f}%
\bm{\mu}_{ge^{\prime}}\bm{\mu}_{e^{\prime}f}\bm{\mu}_{fe}\bm{\mu}%
_{eg}I_{e^{\prime}}^{\ast}\left( t_{1}+t_{2}+t_{3}\right) I_{f}\left(
t_{3}\right) I_{e}\left( t_{2}\right) \text{,}  \notag
\end{align}%
\begin{align}
\bm{S}_{II}^{(SOS)}(t_{3},t_{2},t_{1}) &
=i^{3}\theta(t_{3})\theta(t_{2})\theta(t_{1})\sum_{e,e^{\prime}}\bm{\mu}%
_{ge^{\prime}}\bm{\mu}_{e^{\prime}g}\bm{\mu}_{ge}\bm{\mu}_{eg}I_{e^{\prime}}%
\left( t_{3}\right) I_{e}\left( t_{1}\right)  \label{eigen S2} \\
& +i^{3}\theta(t_{3})\theta(t_{2})\theta(t_{1})\sum_{e,e^{\prime}}\bm{\mu}%
_{ge^{\prime}}\bm{\mu}_{e^{\prime}g}\bm{\mu }_{ge}\bm{\mu}%
_{eg}I_{e^{\prime}}^{\ast}\left( t_{2}\right) I_{e}\left(
t_{1}+t_{2}+t_{3}\right)  \notag \\
& -i^{3}\theta(t_{3})\theta(t_{2})\theta(t_{1})\sum_{e,e^{\prime}}\sum _{f}%
\bm{\mu}_{ge^{\prime}}\bm{\mu}_{e^{\prime}f}\bm{\mu}_{fe}\bm{\mu}%
_{eg}I_{e^{\prime}}^{\ast}\left( t_{2}+t_{3}\right) I_{f}\left( t_{3}\right)
I_{e}\left( t_{1}+t_{2}\right) ,  \notag
\end{align}%
\begin{align}
\bm{S}_{III}^{(SOS)}(t_{3},t_{2},t_{1}) &
=i^{3}\theta(t_{3})\theta(t_{2})\theta(t_{1})\sum_{e,e^{\prime}}\sum_{f}%
\bm{\mu }_{ge^{\prime}}\bm{\mu}_{e^{\prime}f}\bm{\mu}_{fe}\bm{\mu}%
_{eg}I_{e^{\prime}}\left( t_{3}\right) I_{f}\left( t_{2}\right) I_{e}\left(
t_{1}\right)  \label{eigen S3} \\
& -i^{3}\theta(t_{3})\theta(t_{2})\theta(t_{1})\sum_{e,e^{\prime}}\sum _{f}%
\bm{\mu}_{ge^{\prime}}\bm{\mu}_{e^{\prime}f}\bm{\mu}_{fe}\bm{\mu}%
_{eg}I_{e^{\prime}}^{\ast}\left( t_{3}\right) I_{f}\left( t_{2}+t_{3}\right)
I_{e}\left( t_{1}\right) ,  \notag
\end{align}
where $I_{e}(t)$, defined in Eq.~(\ref{I def}), is the Green's function in
the single-exciton eigenstate basis. Eqs.~(\ref{eqn:SOS I freq},\ref{eqn:SOS
II freq},\ref{eqn:SOS III freq}) immediately follow by substituting Eqs.~(%
\ref{eigen S1},\ref{eigen S2},\ref{eigen S3}) in Eq.~(\ref{eqn:pol alpha})
and (\ref{eqn:pol alpha 2}).

\section{Quasiparticle picture for soft-core and hard-core bosons\label%
{App_coup_osc}}

In this Appendix we apply our QP expressions to two other types of
quasiparticles with different statistics. These two examples demonstrate the
generality of our approach discussed briefly in Sec.~\ref{Sec:Discuss}.

We first consider the Hamiltonian of a system of coupled anharmonic
oscillators (soft-core bosons):%
\begin{equation*}
\hat{H}=\sum_{mn}h_{mn}\hat{B}_{m}^{\dagger}\hat{B}_{n}+\sum_{mnkl}U_{mnkl}%
\hat{B}_{m}^{\dagger}\hat{B}_{n}^{\dagger}\hat{B}_{k}\hat{B}_{l}, 
\end{equation*}
where $\hat{B}_{m}^{+}$ and $\hat{B}_{n}$ are boson creation and
annihilation operators with commutation $[B_{m},B_{n}^{+}]=\delta_{mn}$ $%
h_{mm}$ is the fundamental transition energy of the $m$th oscillator, while $%
h_{mn}$ is the coupling between the $m$th and the $n$th oscillators.\ $%
U_{mnkl}$ is the anharmonic coupling. This Hamiltonian has been used to
describe infrared nonlinear spectra of proteins.\cite%
{AbramaviciusMukamel2006b,AbramaviciusMukamel_time2005}

For this model the scattering matrix can be obtained from (\ref%
{eqn:Gamma_Final}) by putting $\mathcal{P}=0.$ In the site representation it
reads:%
\begin{equation*}
\Gamma=\left( \mathbb{I}-V\mathcal{G}_{0}\right) ^{-1}V, 
\end{equation*}
here $\Gamma$ is a tetradic matrix, $V=2U$ and $\mathcal{G}_{0}\left(
\omega\right) $ is defined in Eq.~(\ref{eqn:Bethe}).

We next turn to electronic excitations in molecular aggregates or crystals
with weakly interacting molecules. These are described using the Frenkel
Exciton Hamiltonian. If the excited-state absorption frequency of each
molecule is well separated from the ground state absorption, the excitations
can be modelled as coupled two-level systems.\cite%
{leegwaterMukamel92,juzeliunasknoester2000} The Hamiltonian is%
\begin{equation*}
\hat{H}=\sum_{mn}h_{mn}\hat{B}_{m}^{\dagger}\hat{B}_{n}\text{.}
\end{equation*}
The nonlinearities are now hidden in the statistics of exciton creation $(%
\hat{B}_{m}^{+})$ and annihilation $(\hat{B}_{n})$ operators. These are
bosonic for different oscillators (units) and fermionic for the same
oscillator. Their Pauli commutation relation is $[\hat{B}_{m},\hat{B}%
_{n}^{+}]=\delta_{mn}\left( 1-2\hat{B}_{n}^{\dagger}\hat{B}_{n}\right) $.\
The commutation relation ensures that two excitations are not allowed to
reside on the same site (hard-core bosons). The scattering matrix in this
case is given by:%
\begin{align*}
\Gamma_{mnkl} & =\delta_{mn}\delta_{kl}\bar{\Gamma}_{mn}~, \\
\bar{\Gamma} & =-\mathcal{\bar{G}}\left( \omega\right) ^{-1}~,
\end{align*}
and%
\begin{equation*}
\mathcal{\bar{G}}_{mn}\left( \omega\right) =\delta_{mm_{1}}\delta_{nn_{1}}%
\mathcal{G}_{0mm_{1}nn_{1}}\left( \omega\right) . 
\end{equation*}
This form of the exciton scattering matrix was recently successfully applied
to study molecular chirality induced signals in molecules.\cite{AbraMuk2006}
It can be obtained from Eq.~(\ref{eqn:Gamma_Final}) in the limit $U=0$. All
QP-statistics effects (Paulion commutation relations) are included in the
relation $V_{nmpq}=-2\sum_{l}\mathcal{P}_{nmlp}h_{lq}$ (Eq.~\ref{eqn:anh def}%
) and $\mathcal{P}_{nmlp}=\delta_{nm}\delta_{nl}\delta_{mp}$.


\end{document}